\documentclass[sigconf, nonacm]{acmart}

\newcommand\vldbdoi{10.14778/3430915.3430923}
\newcommand\vldbpages{329 - 341}
\newcommand\vldbpagestyle{empty}
\newcommand\vldbvolume{14}
\newcommand\vldbissue{3}
\newcommand \vldbyear{2021}

\newcommand\vldbauthors{\authors}
\newcommand\vldbtitle{\shorttitle} 
\newcommand\vldbavailabilityurl{https://github.com/yiminl18/LOCATER.git}
\usepackage{amsmath,amsfonts}
\usepackage{algorithmic}
\usepackage{textcomp}
\usepackage{xcolor}
\usepackage{graphicx}
\usepackage[ruled,linesnumbered,vlined]{algorithm2e}
\usepackage[font=small,skip=2pt]{caption}
\usepackage{easy-todo}
\makeatletter
\let\c@lofdepth\relax
\let\c@lotdepth\relax
\makeatother
\usepackage{subfigure}
\usepackage{url}
\usepackage{xcolor}
\usepackage{balance}  
\newtheorem{mydef}{Definition}
\newtheorem{mytheory}{Theorem}
\graphicspath{{images/}}

\newcommand{\squishlist}{
	\begin{list}{$\bullet$}
		{
			\setlength{\itemsep}{0pt}
			\setlength{\parsep}{1pt}
			\setlength{\topsep}{1pt}
			\setlength{\partopsep}{0pt}
			\setlength{\leftmargin}{1.5em}
			\setlength{\labelwidth}{1em}
			\setlength{\labelsep}{0.5em} } }
	
\newcommand{\squishend}{\end{list}}

\begin{document}
\title{LOCATER: Cleaning WiFi Connectivity Datasets for Semantic Localization}


\author{ 
    Yiming Lin, Daokun Jiang, Roberto Yus, Georgios Bouloukakis, \\ Andrew Chio, Sharad Mehrotra, Nalini Venkatasubramanian}
\affiliation{University of California, Irvine, USA.\\ \{yiminl18,daokunj,ryuspeir,gboulouk,achio\}@uci.edu, \{sharad,nalini\}@ics.uci.edu}

\renewcommand{\shortauthors}{Lin et al.}

\begin{abstract}
This paper explores the data cleaning challenges that arise in using WiFi connectivity data to locate users to semantic indoor locations such as buildings, regions, rooms. 
WiFi connectivity data consists of sporadic connections between devices and nearby WiFi access points (APs), each of which may cover a relatively large area within a building. Our system, entitled semantic LOCATion cleanER (LOCATER), postulates semantic localization as a series of data cleaning tasks - first, it treats the problem of determining the AP to which a device is connected between any two of its connection events as a missing value detection and repair problem. 
It then associates the device with the semantic subregion (e.g., a conference room in the region) by postulating it as a location disambiguation problem. LOCATER uses a bootstrapping semi-supervised learning method for  coarse localization and a probabilistic method to achieve finer localization. The paper shows that LOCATER can achieve significantly high accuracy  at both the coarse and fine levels.
\end{abstract}

\maketitle

\pagestyle{\vldbpagestyle}
\begingroup\small\noindent\raggedright\textbf{PVLDB Reference Format:}\\
\vldbauthors. \vldbtitle. PVLDB, \vldbvolume(\vldbissue): \vldbpages, \vldbyear.\\
\href{https://doi.org/\vldbdoi}{doi:\vldbdoi}
\endgroup
\begingroup
\renewcommand\thefootnote{}\footnote{\noindent
This work is licensed under the Creative Commons BY-NC-ND 4.0 International License. Visit \url{https://creativecommons.org/licenses/by-nc-nd/4.0/} to view a copy of this license. For any use beyond those covered by this license, obtain permission by emailing \href{mailto:info@vldb.org}{info@vldb.org}. Copyright is held by the owner/author(s). Publication rights licensed to the VLDB Endowment. \\
\raggedright Proceedings of the VLDB Endowment, Vol. \vldbvolume, No. \vldbissue\ %
ISSN 2150-8097. \\
\href{https://doi.org/\vldbdoi}{doi:\vldbdoi} \\
}\addtocounter{footnote}{-1}\endgroup

\ifdefempty{\vldbavailabilityurl}{}{
\vspace{.3cm}
\begingroup\small\noindent\raggedright\textbf{PVLDB Artifact Availability:}\\
The source code, data, and/or other artifacts have been made available at \url{\vldbavailabilityurl}.
\endgroup
}

\begin{figure*}
	\centering
	\vspace{-0.5em}
	\includegraphics[width=1\linewidth]{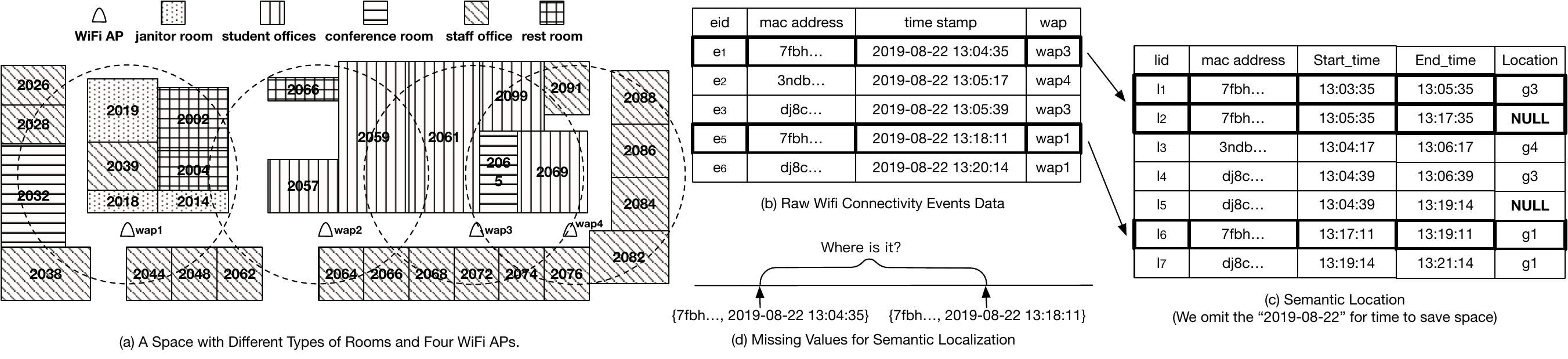}
	\vspace{-1.3em}
	\caption{Motivating Example.}
	\vspace{-1.5em}
	\label{fig:motivating}
\end{figure*}

\section{Introduction}
\label{sec:intro}

This paper studies the challenge of cleaning connectivity data collected by WiFi infrastructures to support {\em semantic localization} inside buildings.
By semantic localization we refer to the problem of {\bf associating a person’s location to a semantically meaningful spatial extent such as a floor, region, or a room.}


Semantic localization differs from (and complements) the well-studied problem of indoor positioning/localization~\cite{priyantha2000cricket,deak2012survey}  that aims to determine the exact physical position of people inside buildings (e.g., coordinate (x,y) within radius r, with z\% certainty). 
If indoor positioning/physical localization could be solved accurately, it would be simple to exploit knowledge about the building’s floor plan and layout to determine the semantic location of the device. However, despite over two decades of work in the area~\cite{liu2007survey,deak2012survey,zafari2019survey}, and significant technological progress, accurate indoor positioning remains an open problem~\cite{zafari2019survey}. Among others, the reasons for this include technology limitations such as costs associated with the required hardware/software \cite{luo2016pallas,youssef2007challenges,xu2013scpl,seifeldin2012nuzzer}, the intrusive nature and inconvenience of these solutions for users~\cite{kang2014smartpdr,deak2012survey,priyantha2000cricket} (who require specialized hardware/software), and algorithmic limitations to deal with dynamic situations such as occlusions, signal attenuation, interference \cite{musa2012tracking,want1992active,li2015passive}. As a result, applications that depend upon accurate positioning and  those that could benefit from semantic localization have faced challenges in effectively utilizing indoor localization technologies.

While indoor localization methods have targeted applications such as 
indoor navigation and augmented reality that require highly accurate  positioning, 
semantic localization suffices for a broad class of smart space applications such as determining occupancy of rooms, thermal control based on occupancy~\cite{afram2014theory}, determining density of people in a space and areas/regions of high traffic in buildings –-applications that have recently gained significance for COVID-19 prevention and monitoring in workplaces~\cite{trivedi2020wifitrace,gupta2020quest}, or locating individuals inside large buildings~\cite{jensen2009graph,musa2012tracking}. 
Despite the utility of semantic localization, to the  best of our knowledge, semantic localization has never before been  studied as a problem in itself.\footnote{Prior papers on indoor localization ~\cite{jia2015soundloc,jiang2012ariel} have evaluated their positioning techniques by measuring
the accuracy at which devices  can be located physically   inside/outside a room. Such work has neither formulated nor addressed the semantic localization challenge  explicitly.  
Instead, 
naive strategies such as degree of spatial overlap/random selection of an overlapping room out of the several choices are used for their experimental study.}

This paper proposes a location cleaning system, entitled {\em LOCATER} to address the  problem of semantic localization. LOCATER can be viewed as a system, the input to which is a log of coarse/\\inaccurate/incomplete physical locations of people inside the building (that could be the result of any indoor positioning/localization strategy or even the raw logs collected by WiFi APs) and the output of which is a clean version of such a log with the semantically meaningful geographical location of the device in the building -- viz., a floor, a region, or, at the fine-granularity, a room. 
Current solutions determine the physical location of a device and use simple heuristics (e.g., largest overlap with the predicted region) for room-level localization. 
In contrast, LOCATER postulates associating a device to a semantic location as a data cleaning challenge and exploits the inherent semantics in the sensor data capturing the building usage to make accurate assessments of device locations. LOCATER, we believe, is the first such system to study semantic localization as a problem in its own right.

While LOCATER could be used alongside any indoor positioning/localization solutions\footnote{See related work for strengths/weaknesses of such technologies.}, we built LOCATER using a localization scheme that uses connectivity events between devices and the WiFi hardware (viz., access points --APs--) that constitute the WiFi infrastructure of any organization.
Such connectivity events, generated in the network when devices connect to an AP, can be collected in real-time using a widely used SNMP (Simple Network Management Protocol), a more recent   NETCONF~\cite{enns2006netconf},  network management protocol, or from  network Syslog~\cite{gerhards2009syslog} containing AP events. Connectivity events consist of observations in the form of $\langle $\textsf{mac address}, \textsf{time stamp}, \textsf{wap}$\rangle$ which correspond to the MAC of the WiFi-enabled connected device, the timestamp when the connection occurred and the WiFi AP (wap) to which the device is connected. Since APs are at fixed locations, connectivity events can be used to locate a device to be in the region covered by the AP.
In Figure~\ref{fig:motivating}(b) an event $e_1$ can lead to the observation that the owner of the the device with mac address \textsf{7bfh...} was located in the region covered by \textsf{wap3} (which includes rooms \textsf{2059}, \textsf{2061}, \textsf{2065}, \textsf{2066}, \textsf{2068}, \textsf{2069}, \textsf{2072}, \textsf{2074}, \textsf{2076}, and \textsf{2099}, in Figure~\ref{fig:motivating}(a)) at \textsf{13:04:35}.

Using WiFi infrastructure for coarse location, as we do in LOCATER,  offers several distinct benefits. First, since it is ubiquitous in modern buildings, using the infrastructure for semantic localization does not incur any additional hardware costs either to users or to the built infrastructure owner. Such would be the case if we were to retrofit buildings with technologies such as RFID, ultra wideband (UWB), bluetooth, camera, etc.~\cite{liu2007survey}. Besides being (almost) zero cost, another artifact of ubiquity of WiFi networks is that such a solution has wide applicability to all types of buildings - airports, residences, office spaces, university campuses, government buildings, etc. Another key advantage is that localization using WiFi connectivity can be performed passively without requiring users to either install new applications on their smartphones, or to actively participate in the localization process.

\noindent
{\bf Challenges in exploiting WiFi connectivity data.}
While WiFi connectivity datasets offer several benefits, they offer coarse localization -- e.g., in a typical office building, a AP may cover a relatively large region consisting of dozens of rooms, and as such, connectivity information does not suffice to build applications that need semantic localization. Using WiFi connectivity data for semantic localization, raises the following technical challenges:

$\bullet$ {\em Missing value detection and repair.} Devices might get disconnected from the network even when the users carrying them are still within the space. Depending on the specific device, connectivity events might occur only sporadically and at different periodicity, making prediction more complex. These lead to a \textit{missing values} challenge. As an example, in Figure~\ref{fig:motivating}(c) we have raw connectivity data for device \textsf{7fbh} at time \textsf{13:04:35} and \textsf{13:18:11}. Location information between these two consecutive time stamps is missing. 

$\bullet$ {\em Location disambiguation.} APs cover large regions within a building that might involve multiple rooms and hence simply knowing which AP a device is connected to may not offer room-level localization. For example, in Figure~\ref{fig:motivating}, the device \textsf{3ndb} connects to \textsf{wap2}, which covers rooms: 2004, 2057, 2059,..., 2068. These values are \textit{dirty} for room-level localization. Such a challenge can be viewed as a location disambiguation challenge. 

$\bullet${\em Scalability.} The volume of WiFi data can be very large - for instance, in our campus, with over 200 buildings and 2,000 plus APs, we generate several million WiFi connectivity tuples in one day on average. Thus, data cleaning technique needs to be able to scale to large data sets. 

To address the above challenges, LOCATER 
uses an iterative classification method that leverages temporal features in the WiFi connectivity data to repair the missing values. Then, spatial and temporal relationships between entities are used in a probabilistic model to disambiguate the possible rooms in which the device may be. LOCATER cleans the WiFi connectivity data in a dynamic setting where we clean objects on demand in the context of queries. 
 In addition, LOCATER caches cleaning results of past queries to speed up the system. 
 Specifically, we make the following contributions:
 (1) We propose a novel approach to semantic indoor localization by formalizing the challenge as a combination of missing value cleaning and disambiguation problems (Section 2)
(2) We propose an iterative classification method to resolve the missing value problem (Section 3) and a novel probability-based approach to disambiguate room locations without using labeled data (Section 4)
(3) We design an efficient caching technique to enable LOCATER to answer queries in near real-time (Section 5)
(4) We validate our approach in a real world testbed and deployment. Experimental results show that LOCATER achieves high accuracy and good scalability on both real and simulated data sets (Section 6). 

\vspace{-1em}


\vspace{0.2em}
\section{Semantic Localization Problem}
\label{sec:definition}
\vspace{-0.3em}
The problem of semantic localization consists of associating for each device its location at any instance of time at a given level of spatial granularity. 

\newcommand{\vBuildings}[1][]{B_{#1}}
\newcommand{\vBuilding}[1][]{b_{#1}}
\newcommand{\vRegions}[1][]{G_{#1}}
\newcommand{\vRegion}[1][j]{g_{#1}}
\newcommand{\vRooms}[1][]{R_{#1}}
\newcommand{\vRoom}[1][j]{r_{#1}}

\newcommand{\vRoomPublic}[1][pb]{\vRooms^{#1}}
\newcommand{\vRoomPrivate}[1][pr]{\vRooms^{#1}}

\newcommand{\vRegionsOfBuilding}[1][]{\vRegions(\vBuilding[{#1}])}
\newcommand{\vRoomsOfRegion}[1][j]{\vRooms(\vRegion[#1])}
\newcommand{\vRoomsOfRegionTime}[1][\vRegion,t]{\vRooms({#1})}

\newcommand{\vPreferRoom}[1][]{R^{pf}(d_{#1})}
\newcommand{\vSystem}[1][]{Safari}

\newcommand{\vDevices}[1][]{D_{#1}}
\newcommand{\vDevice}[1][i]{d_{#1}}
\newcommand{\vNeighborsOfDevice}[1][]{N({\vDevice[{#1}]})}

\newcommand{\vSetLocationOfDevice}[1][]{L{(d_i, {#1}, t)}}
\newcommand{\vLocationOfDevice}[1][]{l{(d_i, {#1}, t)}}

\newcommand{\vAPs}[1][]{WAP_{#1}}
\newcommand{\vAP}[1][j]{wap_{#1}}

\newcommand{\vConEvent}[1][n]{e_{{#1}}}
\newcommand{\vConEvents}[1][\vDevice]{E({#1})}
\newcommand{\vRateConEvents}[1][]{\lambda^{\vConEvents}}

\newcommand{\vGapOfDevice}[1][t_{n},t_{n+1}]{gap_{#1}}
\newcommand{\vGapsOfDevice}[1][]{GAP(\vDevice)}
\newcommand{\vGapsOfDeviceForT}[1][]{GAP_T(\vDevice)}

\newcommand{\vTimeInterval}[1][i]{\delta(\vDevice[{#1}])}
\newcommand{\vTimeIntervalPrime}[1][i]{\delta^{'}(\vDevice[{#1}])}

\newcommand{\vAffinity}[1][]{\alpha({#1})}
\newcommand{\vAffinityDaily}[1][]{\alpha^{day}({#1})}
\newcommand{\vAffinityWeekly}[1][]{\alpha^{week}({#1})}
\newcommand{\vAffinityMonthly}[1][]{\alpha^{month}({#1})}
\newcommand{\vAffinityYearly}[1][]{\alpha^{year}({#1})}
\newcommand{\vProbability}[1][]{p({#1})}
\newcommand{\vWeightRoom}[1][]{w^{pr}({#1})}
\newcommand{\vWeightSocial}[1][]{w^{sc}({#1})}

\newcommand{\vLowBuild}{\tau_l}
\newcommand{\vHighBuild}{\tau_h}
\newcommand{\vLowRegion}{\vLowBuild'}
\newcommand{\vHighRegion}{\vHighBuild'}

\vspace{-0.7em}

\subsection{Space Model}
\vspace{-0.3em}
LOCATER models space  at three levels of spatial granularity\footnote{
The technique can be easily adapted to other spatial models  conforming to the nature of the underlying space.}:

\vspace{0.03in}
\noindent \textit{\underline{Building}}: The coarsest building granularity $B$ takes the values $B = B_1, ..B_n, b_{out}$, where $B_i= 1...n$ represents the set of buildings and $b_{out}$ represents the fact that the device is not in any of the buildings.  We call a device inside a building as \textit{online} device and outside as \textit{offline} device.

\vspace{0.03in}
\noindent \textit{\underline{Region}}: Each building $B_i$ contains a set of regions $G= \{\vRegion : j \in [1 ... |\vRegions|] \}$\footnote{We drop the parameter from $G(B_i)$ and simply refer to it as $G$ since we are dealing with inside a given building.}. We consider a region $\vRegion$ to be the area covered by the network connectivity of a specific WiFi AP~\cite{tervonen2016applying} (represented with dotted lines in Figure~\ref{fig:motivating}(a)). Let $\vAPs = \{\vAP : j \in [1 ... |\vAPs|] \}$ be the set of APs within the building. Hence, $|\vRegions|=|\vAPs|$ and each $\vAP$ is related to one and only one $\vRegion$. Interchangeably, we denote by $Cov(wap_j)$ as the region covered by $wap_j$. 
In Figure~\ref{fig:motivating}(a), there exist four APs $wap_1,..., wap_4$ and thus there exist four regions such that $G=\{g_{1},g_{2},g_{3},g_{4}\}$. Regions can/often do overlap. 

\vspace{0.03in}
\noindent \textit{\underline{Room}}: A building contains a set of rooms $\vRooms = \{\vRoom[j] : [1 ... |\vRooms|]\}$ where $\vRoom[j]$ represents the ID of a room within the building -- e.g., $\vRoom[1] \rightarrow 2065$. Furthermore, a region $\vRegion[i]$ contains a subset of $\vRooms$. 
Let  $\vRoomsOfRegion[i] = \{\vRoom[j] : [1 ... |\vRoomsOfRegion[i]|]\}$ be the set of rooms covered by region $g_i$. Since regions can overlap, a specific room can be part of different regions if its extent intersects with multiple regions. For instance, in Figure~\ref{fig:motivating}(a) room 2059 belongs to both regions $g_{2}$ and $g_{3}$.

\begin{table}[bt]
\vspace{-0.2em}
	\caption{Model variables and shorthand notation.} 
	\centering {\renewcommand{\arraystretch}{1.4} {\scriptsize  
			\begin{tabular}{| p{3cm} | p{4.5cm} |}\hline
				\textbf{Variable(s)}&\textbf{Definition/Description}\\ \hline 
				
				$\vBuildings = \{ B_1,..,B_n, b_{out}\}$; $\vRegion \in \vRegions$; $\vRoom \in \vRooms$ & buildings; regions; rooms\\\hline 
				
				 $\vRoomsOfRegion$ & set of rooms in region $\vRegion$\\\hline

				$\vAP \in \vAPs$; $\vDevice \in \vDevices$ & WiFi APs; devices\\\hline

					$\vTimeInterval$; $gap_{t_{s},t_{e}}(d_i)$  & time interval validity of $d_i;$ gap associated to $\vDevice$ in $[t_{s},t_{e}]$ \\\hline
					
				$l_i \in L$ & semantic location relation \\ \hline
				
			\end{tabular}}}
			\label{table:notations}
			\vspace{-1em}
\end{table}

We consider that rooms in a building have metadata associated. In particular, we classify rooms into two types: (i)~\emph{public}: shared facilities such as meeting rooms, lounges, kitchens, food courts, etc., that are accessible to multiple users (denoted by  $\vRoomPublic \subseteq \vRooms$); and (ii)~\emph{private}: rooms typically restricted to or owned by certain users such as a person's office (denoted by $\vRoomPrivate \subseteq \vRooms$ such that $\vRooms = \vRoomPublic \cup \vRoomPrivate$). 

\vspace{-0.5em}
\subsection{WiFi Connectivity Data} 
Let $\vDevices = \{d_j : j \in [1 ... |\vDevices|] \}$ be the set of devices and $TS=\{t_j : j \in [1 ... |TS|]\}$ the set of time stamps.\footnote{The granularity of $t_j$ can be set on various scenarios.} 
Let $E = \{\vConEvent[i]: i \in [1 ... |E|]\}$ be the WiFi connectivity events table with attributes $\{eid, dev, time, w\}$ corresponding to the event id, device id ($dev \in D$), the time stamp when it occurred ($time \in TS$), and the WiFi AP that generated the event ($w\in WAP$). (As shown in  Figure~\ref{fig:motivating}(b)) 
For each tuple $e_i \in E$, we will refer to each attribute (e.g., $dev$) as $e_i.dev$.\footnote{We use the device's unique MAC address to represent it.}

\begin{figure}[tb]
	\centering
	\vspace{-0.8em}
	\includegraphics[width=1\linewidth]{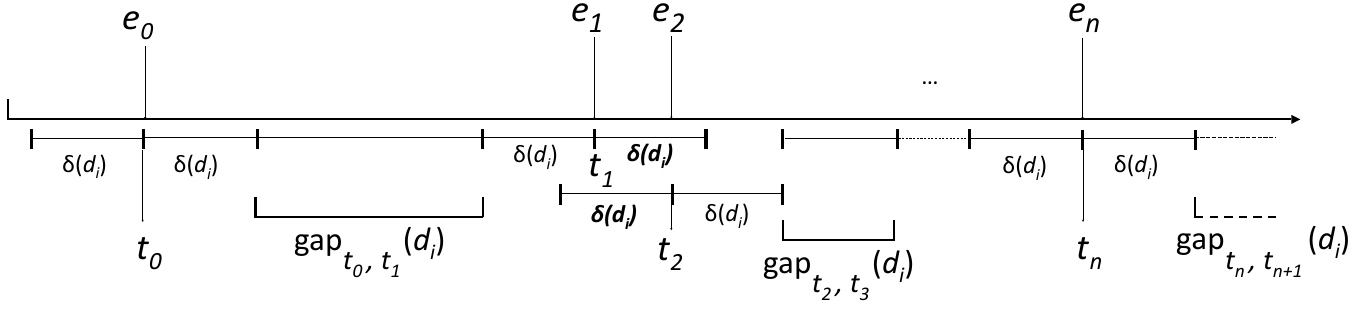}
	\caption{Connectivity events of device $\vDevice[i]$ and their validity.}
	\vspace{-2em}
	\label{fig:con_events}
\end{figure}

Connectivity events occur stochastically even when devices are stationary and/or the signal strength is stable. Events are typically generated when (i)~a device connects to a WiFi AP for the first time, (ii)~the OS of the device decides to probe available WiFi APs around, or (iii)~when the device changes its status. Hence, connectivity logs do not contain an event for every instance of time a device is connected to the WiFi AP or located in a space.
Because of the sporadic nature of connectivity events, we associate to each event a {\em validity} period denoted by $\delta$. 
The value of $\delta$ depends on the actual device $d_i$ (in the extended version of the paper~\cite{locater2020} we show how to estimate $\delta$) and is denoted by $\delta(d_i)$ (see Figure~\ref{fig:con_events} for some sample connectivity events of device $d_i$). 
Let the Valid Interval for an event $e_i$ be $VI_i=\{VI_i.st, VI_i.et\}$, where $VI_i.st$ ($VI_i.et$) is the start (end) time stamp of this interval.   
Considering the connectivity events of device $d_i$, the valid interval for event $e_i$ can be considered in three ways. 1) If the subsequent (previous) event $e_j$ of the same device happens after (before) $e_i.time+\delta(e_i.dev)$ ($e_i.time-\delta(e_i.dev)$), then $VI_i.et = e_i.time+\delta(e_i.dev)$ ($VI_i.st = e_i.time-\delta(e_i.dev)$); (e.g., event $e_0$ in Figure~\ref{fig:con_events}) 2) Otherwise, if the subsequent (previous)  event $e_j$ happens close to $e_i$ ($|e_i.time-e_j.time| < \delta(e_i.dev)$), $VI_i.et = e_j.time$ ($VI_{i}.st = e_i.time$). (e.g., $e_1$ is valid in $(t_1-\delta(d_i),t_2)$, and $e_2$ is valid in $(t_2,t_2 + \delta(d_i))$ in  Figure~\ref{fig:con_events}). 
While we assume that an event is valid for $\delta$ period, there can be portions of time in which no connectivity event is valid in the log for a specific device. We refer to such time periods as {\em gaps}.
Let $gap_{t_{s},t_{e}}(d_i)$ be the gap of device $d_i$ that starts at $t_{s}$ and ends at $t_{e}$ time stamp. In Fig~\ref{fig:con_events}, $gap_{t_{0},t_{1}}(d_i)$ represents a gap of $d_i$ whose time interval is $[t_{0}, t_{1}]$.

\vspace{-0.6em}
\subsection{Semantic  Location Table} 
\vspace{-0.2em}
The semantic localization challenge (i.e., determining the location of device $d_i$ at any time $t_j$ at a given spatial granularity) can be viewed as 
equivalent to creating  a \textsf{Semantic Location Table}, $L=\{l_i: i\in [1...|L|]\}$,  with the attributes $\{lid, dev, loc, st, et\}$ such that the device $dev$ is in the location $loc$ from time $st$ to $et$. The table $L$ is such that for any device $dev$ and any time $t$, there exists a tuple in $L$ such that $st \leq t \leq et$, (i.e., the  table covers the location of each device at all times under consideration). 

We can form the table $L$ from the event table $E$ as follows:
for each event $e_i \in E$ we create a corresponding tuple $l_j \in L$, where $l_j.dev = e_i.dev$, $l_j.loc = Cov(e_i.w)$, and its start and end times correspond to the validity interval of the event $e_i$, i.e., $l_j.st = VI_i.st$ and $l_j.et = VI_i.et$.
We further insert a tuple $l_j$ corresponding to each gap in the event table $E$. For each gap $gap_{t_{s},t_{e}}(d_i)$, we generate a tuple $l_j \in L$ such that $dev_j=d_i$, $st_j = t_{s}$, $et_j = t_{e}$, $loc_j = $ \textsf{NULL}. 
Furthermore, let $L^{c}=\{l_i : loc_i \neq \textsf{NULL}\}$ be the set of tuples whose location is not NULL, and $L^{d} = L \setminus L^{c}$ be the set of tuples whose  location is \textsf{NULL}. 
We further define  $L(d_j)=\{l_i : dev_i = d_j\}$ as the set of tuples of device $d_j$ and $L_{T}$ be the set of tuples of device $d_i$ happening in time period $T$.

In Fig~\ref{fig:motivating}(c), we transform raw WiFi connectivity data to a semantic location table. In this example, we assume $\delta=1$ minute for all devices. $e_1$ in Fig~\ref{fig:motivating}(b) corresponds to $l_1$ in Fig~\ref{fig:motivating}(c), where time stamp is expanded to a valid interval, and the gap between $e_1$ and $e_5$ in Fig~\ref{fig:motivating}(b) corresponds to the tuple $l_2$ in Fig~\ref{fig:motivating}(c). 

\vspace{-0.6em}
\subsection{Data Cleaning Challenges}
\vspace{-0.2em}
The table $L$, which captures semantic location of individuals, contains two data cleaning challenges corresponding to coarse and fine-grained localization.

\noindent{\em \textbf{ Coarse-Grained Localization}:}
 Given a tuple $l_i$ with $l_i.loc= \textsf{NULL}$, consists of imputing the missing location value to a coarse-level location by replacing it by either $l_i.loc=b_{out}$ or $l_i.loc = g_j$ (for some region $g_j$ in building $B_k$). $\Box$
  
\noindent{\em \textbf{Fine-Grained Localization}:} 
 Given a tuple $l_i$ with $l_i.loc= g_j$, consists of determining the room $r_k \in R(g_j)$ the device $l_i.dev$ is located in and updating $l_i.loc= r_k$.   $\Box$



We can choose to clean the entire relation $L$ or clean it on demand at query time. In practice applications do not require knowing the fine-grained location of all the users at all times. Instead, they pose point queries, denoted by $\textsf{Query}=(d_{i},t_q)$, requesting the location of device $d_i$ at time $t_q$. Hence, we will focus on 
cleaning the location of the tuple of interest at query time.\footnote{Notice that we could use query-time cleaning to clean the entire relation $L$ by iteratively cleaning each tuple, though if the goal is to clean the entire table better/more efficient approaches would be feasible. Such an approach, however, differs from our focus on real-time queries over collected data. Similar query-time approaches have been considered recently in the context of online data cleaning~\cite{giannakopoulou2020cleaning,altwaijry2013query}.} Thus, given a query $(d_i, t_q)$, LOCATER first determines the tuple in $L$ for the device $d_i$ that covers the time $t_q$. If the location specified in the tuple is \textsf{NULL}, the coarse-level localization algorithm is executed to determine first the region in which the device is expected to be. If fine-grained location is required, the  fine-grained localization algorithm is executed to disambiguate amongst the rooms in the region.

\section{Coarse-Grained Localization}
\label{sec:coarse}

LOCATER uses an iterative classification algorithm combined with bootstrapping techniques to fill in the missing location of a tuple $l_m$ with $l_m.loc= \textsf{NULL}$ for device $l_m.dev$ (in the following we will refer to such tuple as a {\em dirty} tuple).
For simplicity, we use $dev_i, st_i, et_i$ and $loc_i$ to denote $l_i.dev,  l_i.st$, $l_i.et$, $l_i.loc$, respectively. 

The algorithm takes as input, $L_{T}(dev_m)$, a set of historical tuples of device $dev_m$ in time period $T$ consisting of $N$ past days before query time, where $N$ is a parameter set experimentally (see Section~\ref{sec:evaluation}).
For a tuple $l_i$, let $st_i.time$ ($st_i.date$) be the time (date) part of the start timestamp, similarly for $ed_i.time$ ($ed_i.date$).  
Likewise, let $st_i.day$ (and $et_i.day$) refer to the day of the week.\footnote{We assume that gaps do not span multiple days.}
We define the following features for each tuple $l_i \in L_{T}(dev_i)$: 
\squishlist
	\item 
	$st_i.time$, $et_i.time$: the start and end time of tuple $l_j$. 
	\item {\em duration} $\delta(l_j)$: the duration of the tuple (i.e., $et_i.time - st_i.time$). 
	\item  $st_i.day$ ($et_i.day$): the day of the week in which tuple $l_j$ occurred (ended). 
	\item  $loc_{i-1}$, $loc_{i+1}$: the associated region at the start and end time of the tuple. 
	\item {\em connection density} $\omega$: the average number of logged connectivity events (clean tuples) for the device $dev_i$ during the same time period of $l_i$ for each day in $T$.
\squishend

The iterative classification method trains two logistic regression classifiers based on such vectors to label gaps as: 1) Inside/outside and 2) Within a specific region, if inside.

\vspace{0.1cm}
\noindent
\textbf{Bootstrapping.} The bootstrapping process labels a dirty tuple as inside or outside the building by using heuristics that take into consideration the duration of the dirty tuple (short duration inside and long duration outside). We set two thresholds, $\vLowBuild$ and $\vHighBuild$, such that a tuple is labeled as $b_{in}$ if $\delta(l_j) \leq \vLowBuild$ and as $b_{out}$ if $\delta(l_j) \geq \vHighBuild$ (we show two methods to compute $\vLowBuild$ and $\vHighBuild$ in Section~\ref{sec:appendix}). If the duration of a tuple is between $\vLowBuild$ and $\vHighBuild$, then we cannot label it as either inside/outside using the above heuristic. Such dirty tuples are marked as {\em unlabeled}. We partition the set of dirty tuples of device $d_i$,  $L^{d}_{T}(dev_m)$, into two subsets -- $\mathcal{S}_{labeled}$, $\mathcal{S}_{unlabeled}$. 
For tuples in $\mathcal{S}_{labeled}$ that are classified as inside of the building, to further label them with a region at which the device is located, the heuristic takes into account the start and end region of the gap as follows:

\squishlist
    \item 
    If $loc_{j-1}=loc_{j+1}$, then the assigned label is $loc_{j-1}$ (i.e., if the regions at the start and end of the tuple are the same, the device is considered to be in the region for the entire duration).
    \item Otherwise, 
    we assign as label a region $g_k$ which corresponds to the most visited region of $dev_j$ in  connectivity events that overlap with the dirty tuple (i.e., whose connection time is between $st_j.time$ and $et_j.time$).
\squishend

\setlength{\textfloatsep}{0pt}
\begin{algorithm}[bt]
    \small
	\caption{Iterative classification algorithm.}
	\label{alg:semi}
	\KwIn{$\mathcal{S}_{labeled},\mathcal{S}_{unlabeled}$}  
	\While{$\mathcal{S}_{unlabeled}$ is not empty}
	{$ classifier \leftarrow$ TrainClassifier($\mathcal{S}_{labeled}$)\;
	$max\_confidence \leftarrow -1, candidate\_tuple \leftarrow NULL$\;
	\For{$tuple \in \mathcal{S}_{unlabeled}$}{
	    $prediction\_array, label \leftarrow Predict(classifier, tuple)$\;
	    $confidence \leftarrow variance(prediction\_array)$\;
	    \If{$confidence > max\_confidence$}{
	        $max\_confidence \leftarrow confidence$\;
	        $candidate\_tuple \leftarrow tuple$\;
	     }
	}
	$\mathcal{S}_{unlabeled} \leftarrow \mathcal{S}_{unlabeled} - candidate\_tuple$ \;
	$\mathcal{S}_{labeled} \leftarrow \mathcal{S}_{labeled} + (candidate\_tuple,label)$ \;
	}
	\textbf{return} $classifier$; 
\end{algorithm}

\noindent
\textbf{Iterative Classification.} We use iterative classification to label the remaining (unlabeled) dirty tuples $\mathcal{S}_{unlabeled}$, as described in Algorithm~\ref{alg:semi}. For each device $d_i$, we learn logistic regression classifiers on $\mathcal{S}_{labeled}$ (function \texttt{TrainClassifier}$(\mathcal{S}_{labeled})$ in Algorithm~\ref{alg:semi}), which are then used to classify the unlabeled dirty tuples associated with the device.\footnote{We assume that connectivity events exist for the device in the historical data considered, as is the case with our data set. If data for the device does not exist, e.g., if a person enters the building for the first time, then, we can label such devices based on aggregated location, e.g., most common label for other devices.}

Algorithm~\ref{alg:semi} is firstly executed at building level to learn a model to classify if an unlabeled dirty tuple is inside/outside the building. To this end, let $\mathcal{L}$ be the set of possible training labels - i.e., inside/outside the building. The method \texttt{Predict}$(classifier, gap)$, returns an array of numbers from 0 to 1, where each number represents the probability of the dirty tuple being assigned to a label in $\mathcal{L}$ (all numbers in the array sum up to 1), and the label with highest probability in the array. In the array returned by \texttt{Predict}, a larger variance means that the probability of assigning a certain label to this dirty tuple is higher than other dirty tuples. Thus, we use the variance of the array as the confidence value of each prediction. 
In each outer iteration of the loop (lines 1-11), as a first step, a logistic regression classifier is trained on $\mathcal{S}_{labeled}$. Then, it is applied to all tuples in $\mathcal{S}_{unlabeled}$. For each iteration, the dirty tuple with the highest prediction confidence is removed from $\mathcal{S}_{unlabeled}$ and added to $\mathcal{S}_{labeled}$ along with its predicted label. This algorithm terminates when $\mathcal{S}_{unlabeled}$ is empty and the classifier trained in the last round will be returned. 
The same process is followed to learn a model at the region level for dirty tuples labeled as inside the building. In this case, when executing the algorithm $\mathcal{L}$ contains the set regions in the building (i.e., $G$). The output is a classifier that labels a dirty tuple with the region where the device is located.

Given the two trained classifiers, for a dirty tuple $l_m$, we first use the inside/outside classifier to classify $l_m$ as inside or outside of the building. If the tuple $l_m$ is classified as outside, then $loc_m=b_{out}$. Otherwise, we further classify the tuple $l_m$ using the region classifier to obtain its associated region. Then, the device will be located in such region and LOCATER will perform the room-level fine-grained localization as we will explain in the following section.

\vspace{-0.5em}
\section{Fine-Grained Localization}
\label{sec:fine}
\vspace{-0.2em}
Given a query $Q=(d_i,t_q)$ and the associated tuple $l_m$ whose location has been cleaned by the coarse-level localization algorithm, this step determines the specific room $r_j\in R(l_m.loc)$ where $d_i$ is located at time $t_q$.
As shown in Figure~\ref{fig:motivating}(c), tuples $l_1, l_3$, are logged for two devices $d_1$ and $d_2$ with MAC addresses \textsf{7fbh} and \textsf{3ndb}, respectively. 
Assume that we aim to identify the room in which device $d_1$ was located at \textsf{2019-08-22 13:04}. 
Given that $d_1$ was connected to \textsf{wap3} at that time, the device should have been located in one of the rooms in that region $\vRegion[3]$ -- i.e., $\vRoomsOfRegion[3]=\{\textsf{2059},\textsf{2061},\textsf{2065},\textsf{2069},\textsf{2099}\}$. These are called \textit{candidate rooms} of $d_1$ (we omit the remaining candidate rooms -- $\textsf{2066}$, $\textsf{2068}$, $\textsf{2072}$, and $\textsf{2074}$ -- for simplicity). 
The main goal of the fine-grained location approach, is to identify in which candidate room $d_1$ was located.

\vspace{0.1cm}
\noindent
\textbf{Affinity.} LOCATER's location prediction is based on the concept of \emph{affinity} which models relationships between devices and rooms. 
\squishlist
\item \emph{Room affinity}:  $\vAffinity[{\vDevice[i],\vRoom[j],t_q}]$ denotes the affinity between a device $\vDevice[i]$ and a room $\vRoom$ (i.e., the chance of $\vDevice[i]$ being located in $\vRoom[j]$ at time $t_q$), given  the region $g_{k}$ in which $d_i$ is located at time $t_q$. 

\item \emph{Group affinity}:  $\alpha(D,\vRoom[j],t_q)$ represents the affinity of a set of devices $D$ to be in a room $\vRoom[j]$ at time $t_q$ (i.e., the chance of all devices in $D$ being located in $\vRoom[j]$ at $t_q$), given that device $d_i \in D$ is located in region $g_{k}$ at time $t_q$. 
\squishend

\begin{figure}[tb]
	\centering
	\includegraphics[width=0.99\linewidth]{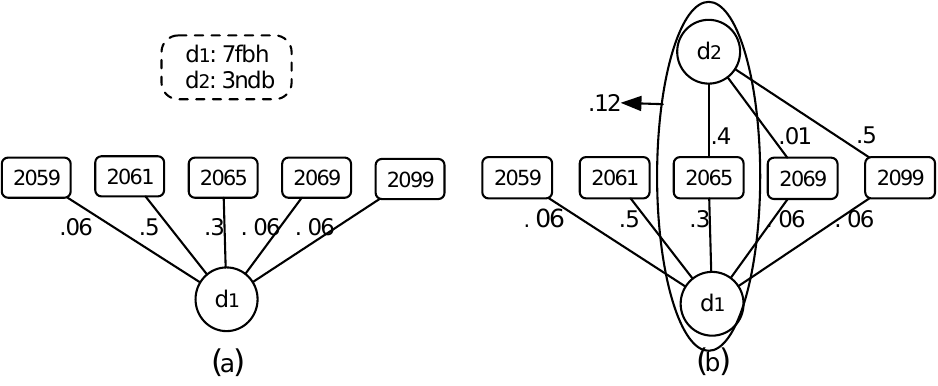}
	\caption{Graph view in fine-grained location cleaning.}
	\vspace{-0.2em}
	\label{fig:graph}
\end{figure}

Note that the concept of group affinity generalizes that of room affinity. While room affinity is a device's conditional probability of being in a specific room, given the region it is located in, group affinity of a set of devices represents the probability of the the set of devices being co-located in a specific room $r_j$ at $t_q$. We differentiate between these since the methods we use to learn these affinities are different, as will be discussed in the following section. We first illustrate how affinities affect localization prediction using the example in Figure~\ref{fig:graph}, which shows a hypergraph representing room and group affinities at time $t_q$. For instance, an edge between $d_1$ and the room $2065$ shows the affinity $\alpha(d_1, 2065, t_q) = 0.3$. 
Likewise the hyperedge $\langle d1, d2, 2065\rangle$ with the label 0.12 represents the group affinity, represented as $\alpha(\{d_1,d_2\},2065,t_q)=0.12$.
If at time $t_q$ device $d_2$ is not online (i.e., there are no events associated with $d_2$ at $t_q$ in that region), we can predict that $d_1$ is in room \textsf{2061} since $d_1$'s affinity to \textsf{2061} is the highest. On the other hand, if $d_2$ is online at $t_q$, the chance that $d_1$ is in room \textsf{2065} increases due to the group affinity $\alpha(\{d_1,d_2\},2065,t_q)=0.12$. 
The location prediction for a device $d_i$, thus, must account for both {\em room} and {\em group affinity}. 

\vspace{0.1cm}
\noindent
\textbf{Room Probability.} Let $Pr(d_{i},r_{j},t_q)$ be the probability that a device $\vDevice[i]$ is in room $\vRoom[j]$ at time $t_q$. Given a query $Q=(\vDevice, t_q)$ and its associated tuple $l_m$, the goal of the fine-grained location prediction algorithm is to find the room $r_j\in R(l_m.loc)$ of $\vDevice$ at time $t_q$, such that $r_j$ has the maximum $Pr(d_{i},r_{j},t_q),  \forall \vRoom\in r_j\in R(l_m.loc)$.
We develop such an algorithm based on estimating $Pr(d_{i},r_{j},t_q)$ based on both room and group affinities in Section~\ref{sec:appendix}. 
Before we discuss the algorithm, we first describe how affinities are estimated. 
 
\vspace{-0.7em}
\subsection{Affinity Learning}
\label{subsec:affinity}
\vspace{-0.1em}

\noindent\textbf{Learning Room Affinity.}
One of the challenges in estimating room affinity is the potential lack of historical room-level location data for devices - collecting such a data would be prohibitively expensive, specially when we consider large spaces with tens of thousands of people/devices. Our approach, thus, does not assume availability of room-level localization data which could have been used to train specific models.\footnote{Extending our approach to handle when such data is obtainable for at least a subset of devices (e.g., through crowd-sourcing) is interesting and part of our future work.} Instead, we compute it based on the available background knowledge and space metadata. 

To compute $\vAffinity[{\vDevice[i],\vRoom[j],t_q}]$, we associate for each device $\vDevice[i]$ a set of preferred rooms $\vPreferRoom[i]$ -- e.g., the personal room of  $\vDevice[i]$'s owner (space metadata), or the most frequent rooms $\vDevice[i]$'s owner enters (background knowledge). $\vPreferRoom[i]$ is an empty set if $\vDevice[i]$'s owner does not have any preferred rooms. 
If $\vRoom[j]$ is one the preferred rooms of $\vDevice[i]$ ($\vRoom[j] \in \vPreferRoom[i]$), we assign to $\vRoom[j]$ the highest weight denoted by $w^{pf}$. Similarly, if $\vRoom[j]$ is a public room ($\vRoom[j]\in (\vRoomsOfRegion[x]\cap \vRoomPublic)\diagdown \vPreferRoom[i]$), we assign to $\vRoom[j]$ the second highest weight denoted by $w^{pb}$. Finally, if $\vRoom[j]$ is a private room ($\vRoom[j]\in (\vRoomsOfRegion[x]\cap \vRoomPrivate)\diagdown \vPreferRoom[i]$), we assign to $\vRoom[j]$ the lowest weight denoted by $w^{pr}$. In general, these weights are assigned based on the following conditions: (1)~$w^{pf}>w^{pb}>w^{pr}$ and (2) $w^{pf}+w^{pb}+w^{pr}=1$. 
The influence of different combinations of $w^{pf},w^{pb},w^{pr}$ is evaluated in Section~\ref{sec:evaluation}. 

We illustrate the assignment of these weights by using the graph of our running example. As already pointed out, $\vDevice[1]$ connects to \textsf{wap3} of region $\vRegion[3]$, where 
$\vRoomsOfRegion[3]=\{\textsf{2059},$ $\textsf{2061},$ $\textsf{2065},$ $\textsf{2069},$ $\textsf{2099}\}$.
In addition, $\vDevice[1]$'s office, room $\textsf{2061}$, is the only preferred room ($\vPreferRoom[1]=\{\textsf{2061}\}$) and $\textsf{2065}$ is a public room (meeting room). Hence, the remaining rooms in $\vPreferRoom[1]$ are other personal offices associated with other devices. Based on Figure~\ref{fig:graph}, a possible assignment of  $w^{pf},w^{pb},w^{pr}$ to the corresponding rooms is as follows:
$\alpha(d_1,\textsf{2061},t_q) = \frac{w^{pf}}{1}=0.5$, $\alpha(d_1,\textsf{2065},t_q)=\frac{w^{pb}}{1}=0.3$, and any room in $\vRoomsOfRegion[3]\setminus (\vPreferRoom[1] \cup \vRoomPublic$) -- 
i.e., $\{\textsf{2059},$ $\textsf{2069},$ $\textsf{2099}\}$ shares the same room affinity, which is $\frac{w^{pr}}{3}=0.066$.

Note that since room affinity is not data dependent, we can pre-compute and store it to speed up computation. Furthermore, preferred rooms could be time dependent (e.g., user is expected to be in the break room during lunch, while being in office during other times). Such a time dependent model   would potentially result in more accurate room level localization if such metadata is available.

\vspace{0.1cm}
\noindent

\noindent\textbf{Learning Group Affinity.} 
\label{subsec:group-affinity}
Before describing how we compute group affinity, we first define the concept of \emph{device affinity}, denoted by $\alpha(D)$, which intuitively captures the probability of devices/users to be part of a group and be co-located (which serves as a basis to compute group affinity).  
Consider all the tuples in $L$. Let $L(d_i)=\{l_j: dev_j=d_i\}$ be the set of tuples corresponding to device $d_i \in D$, and $L(D)$ be the tuples of devices in $D$. Consider the set of semantic location tuples such that for each tuple $l_a \in L(d_i)$, belonging to that set, and for every other device $d_j \in D\setminus d_i$, there exists a tuple $l_b \in L(d_j)$ where devices $l_a.dev$ and $l_b.dev$ are in the same region at (approximately) the same time, i.e., $TR_a \cup TR_b \neq \varnothing$ and $l_a.loc=l_b.loc$ (not \textsf{NULL}).  Intuitively, such a tuple set, referred to as the intersecting tuple set, represents the times when all the devices in $D$ are in the same area (since they are connected to the same WiFi AP). 
We compute device affinity $\alpha(D)$ as a fraction of such intersecting tuples among all tuples in $L(D)$.

Given device affinity $\alpha(D)$, we can now compute the group affinity among devices $D$ in room $\vRoom[j]$ at time $t_q$, i.e., $\alpha(D,\vRoom[j],t_q)$. Let $R_{is}$ be the set of intersecting rooms of connected regions for each device in $D$ at time $t_q$: $R_{is}=\bigcap R(l_i.loc), l_i\in L_{t_q}(D)$. If $r_j$ is not one of the intersecting rooms, $r_j \notin R_{is}$, then $\alpha(D,\vRoom[j],t_q)=0$. Otherwise, to compute $\alpha(D,\vRoom[j],t_q)$, 
we first determine conditional probability of a device $d_i \in D$ to be in $r_j$ given that $r_j \in R_{is}$ at time $t_q$.

Let $@(d_i,r_j,t_q)$ represent the fact that device $d_i$ is in room $r_j$ at time $t_q$, and likewise $@(d_i,R_{is}, t_q)$ represent the fact that $d_i$ is in one of the rooms in $R_{is}$ at $t_q$.
{\small{$P(@(d_i,r_j,t_q)|@(d_i,R_{is},t_q))$ $=$ $\frac{P(@(d_i,r_j,t_q))}{P(@(d_i,R_{is},t_q))}$}}, where  {\small{$P(@(d_i,R_{is},t_q)) = \sum_{r_k\in R_{is}}P(@(d_i,r_k,t_q))$}}. We now compute  $\alpha(D,\vRoom[j],t_q)$, where 
 $r_j\in R_{is}$ as follows: 

\vspace{-1em}
\begin{equation}
\small
    \alpha(D,\vRoom[j],t_q)=\alpha(D)\prod_{d_i\in D}P(@(d_i,r_j,t_q)|@(d_i,R_{is},t_q))
\end{equation}
\vspace{-1em}

Intuitively,  group affinity captures the probability of the set of devices to be  in a given room (based on the
room level affinity of individual devices) given that the (individuals carrying the) devices are co-located, which is captured using the device affinity.

We explain the notation using the example in Figure~\ref{fig:graph}(b).  
Let us assume that the device affinity between $d_1$ and $d_2$ (not shown in the figure) is $.4$, i.e., $\alpha(\{d_1,d_2\})=.4$. The set $R_{is}=\{\textsf{2065},\textsf{2069},\textsf{2099}\}$. 
We compute $\alpha(\{d_1,d_2\},\textsf{2065},t_q)$ as 
{\small{$P(@(d_1,\textsf{2065},t_q)|@(d_1, R_{is},t_q))=\frac{.3}{.3+.06+.06}=.69$}}. Similarly, {\small{$P(@(d_2,\textsf{2065},t_q)|@(d_2, R_{is},t_q))=\frac{.4}{.4+.01+.5}=.44$}}. Finally, {\small{$\alpha(\{d_1,d_2\},\textsf{2065},t_q)=.4*.69*.44=.12$}}.

\subsection{Localization Algorithm}
\label{subsec:alg}

Given a query $Q=(d_i,t_q)$, its associated tuple $l_m$, and candidate rooms $R(l_m.loc)$, we compute the room probability $Pr(d_i,r_j,t_q)$ for each $r_j \in R(l_m.loc)$ and select the room with highest probability as an answer to $Q$. 
We first define the concept of the set of \textit{neighbor} devices of $d_i$, denoted by $D_{n}(d_i)$.
A device $d_k \in D_n(d_i)$ is a \textit{neighbor} of $d_i$ if: (i)~$d_k$ is online at time $t_q$ (inside the building); (ii)~ $\alpha(\{d_i,d_k\},r_j,t_q)>0$ for each $r_j\in R(l_m.loc)$; and (iii)~$R(l_m.loc)\cap R(g_{y})\neq \varnothing$, where $R(g_{y}$) is the region in which $d_k$ is located. In Figure~\ref{fig:graph}(b), $d_2$ is a neighbor of $d_1$. 
Essentially, neighbors of a device $d_i$ could influence the location prediction of $d_i$ (since they will contribute  a non-zero group affinity for $d_i$).

Since we use the concept of neighbor always in the context of a device $d_i$, we will simplify the notation and refer to $D_n(d_i)$ as $D_n$.
Since processing every device in $D_n$ can be computationally expensive, the localization algorithm considers the neighbors iteratively 
until there is enough confidence that the unprocessed devices will not change the current answer.
Let $\bar{D}_n\subseteq D_n$ be the set of devices that the algorithm has processed. 
We denote as $P(\vRoom[j]|\bar{D}_{n})$ the probability of $\vRoom[j]$ being the answer of $Q$  given the devices 
and their locations in  $\bar{D}_{n}$\footnote{We could express the above, as explained in Section~\ref{subsec:group-affinity}, as 
$P( @(d_i, \vRoom[j], t_q)|\bar{D}_{n})$ but we simplify the notation for brevity of following formulas.  
$r_j$ being the answer of query $Q$ means $d_i$ is in $r_j$ at time $t_q$, and we write $r_j$ here for simplicity.} that have been processed by the algorithm so far. 
Using Bayes's rule:  
\vspace{-0.5em}
\begin{equation}
\label{eq:main}
\small
    P(r_j|\bar{D}_{n}) = \frac{P(\bar{D}_{n}|r_j)P(r_j)}{P(\bar{D}_{n}|r_j)P(r_j)+P(\bar{D}_{n}|\lnot r_j)P(\lnot r_j)}
\end{equation}

\noindent where we estimate $P(r_j)$ using the room affinity $\alpha(d_i,r_j,t_q)$. 

We first compute $P(r_j|\bar{D}_{n})$ under the simplifying assumption that probability of $d_i$ to be in room $r_j$ given any two neighbors in $D_n$ is conditionally independent. Then, we consider that multiple neighbor devices may together influence the probability of $d_i$ to be in room $r_j$.

\begin{figure}
\centering
\subfigure[\small Independent Neighbor Set.]{
\includegraphics[width=0.48\linewidth]{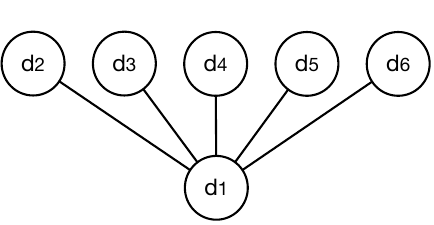}}
\subfigure[\small Dependent Neighbor Set.]{
\includegraphics[width=0.48\linewidth]{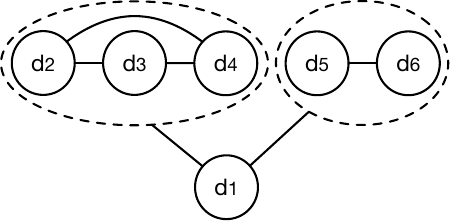}}
\caption{\small Graph view in fine-grained location cleaning.} 
\label{fig:fine-algorithm}
\end{figure}

\vspace{0.03in}
\noindent\textbf{Independence Assumption.}  
Since we have assumed conditional independence:
 $P(\bar{D}_{n}|r_j) = \prod_{d_{k}\in \bar{D}_{n}}P(@(d_k,r_j,t_q)|r_j)$ 
 where\\ $@(d_k,r_j,t_q)$ represents that $d_k$ is located in $r_j$ at time $t_q$. By definition, $P(@(d_k,r_j,t_q)|r_j)=\frac{P(@(d_k,r_j,t_q),r_j)}{P(r_j)}$.
The numerator represents the group affinity, i.e., $P(@(d_k,r_j,t_q),r_j)=\alpha(\{d_k,d_i\},r_j,t_q)$.  
Similarly, $P(@(d_k,r_j,t_q),\neg r_j)=1-\alpha(\{d_k,d_i\},r_j,t_q)$. 
\vspace{-0.2em}
\begin{equation}
\small
    P(r_j|\bar{D}_{n}) = 1/\bigg(1+\frac{\prod_{d_{k}\in \bar{D}_{n}}(1-\alpha(\{d_k,d_i\},r_j,t_q))}{\prod_{d_{k}\in \bar{D}_{n}}\alpha(\{d_k,d_i\},r_j,t_q)}\bigg)
\end{equation}
\vspace{-0.5em}

To guarantee that our algorithm determines the answer of $Q$ by processing the minimum possible number of devices in $\bar{D}_{n}$, we compute the expected/max/min probability of $\vRoom[j]$ being the answer based on neighbor devices in $D_n$.  
We consider the processed devices $\bar{D}_{n}$ as well as unprocessed devices $D_n \setminus \bar{D}_{n}$. Thus, we consider all the possible room locations (given by coarse-location) for unprocessed devices. We denote the set of all possibilities for locations of these devices (i.e., the set of possible worlds~\cite{possible-world}) by $\mathcal{W}(D_{n}\setminus \bar{D}_{n})$. 
For each possible world $W \in \mathcal{W}(D_{n}\setminus \bar{D}_{n})$, let $P(W)$ be the probability of the world $W$ and $P(\vRoom[j]|\bar{D}_{n},W)$ be the probability of $\vRoom[j]$ being the answer of $Q$ given the observations of processed devices $\bar{D}_{n}$ and the possible world $W$. We now formally define the expected/max/min probability of $\vRoom[j]$ given all the possible worlds.

\vspace{-0.5em}
\begin{mydef}
Given a query $Q=(d_i,t_q)$, a region $R(g_x)$, a set of neighbor devices $D_n$, a set of processed devices $\bar{D}_{n} \subseteq D_n$, and the candidate room $r_j \in  R(g_x)$ of $d_i$, the expected probability of $r_j$ being the answer of $Q$, denoted by $expP(r_j|\bar{D}_{n})$, is defined as follows: 
\begin{equation}
\small
    expP(r_j|\bar{D}_{n}) = \sum_{W\in \mathcal{W}(D_{n}\setminus \bar{D}_{n})}P(W)P(r_j|\bar{D}_{n},W)
\end{equation}
\vspace{-0.2em}
The maximum probability of $r_j$, denoted by $maxP(r_j|\bar{D}_{n})$, is:
\begin{equation}
\small
    maxP(r_j|\bar{D}_{n}) = \max\limits_{W\in \mathcal{W}(D_{n}\setminus \bar{D}_{n})}P(r_j|\bar{D}_{n},W)
\end{equation}
\noindent The minimum probability can be defined similarly.
\end{mydef}
\vspace{-0.5em}

The algorithm terminates the iteration only if there exists a room $r_i \in R(g_x)$, for any other room $r_j \in R(g_x), r_i\neq r_j$, such that 
$minP(r_i|\bar{D}_{n})>maxP(r_j|\bar{D}_{n})$. However, it is often difficult to satisfy such strict condition in practice. Thus, we relax this condition using the following two conditions: 
\vspace{-0.2em}
\begin{enumerate}
\small
    \item $minP(r_i|\bar{D}_{n})>expP(r_j|\bar{D}_{n})$(or $P(r_j|\bar{D}_{n})$)
    \item $expP(r_i|\bar{D}_{n})$(or $P(r_i|\bar{D}_{n}))>maxP(r_j|\bar{D}_{n}$)
\end{enumerate}
\vspace{-0.2em}
In Section~\ref{sec:evaluation} we show that these loosen conditions enable the algorithm to terminate efficiently without sacrificing the quality of the results. 

\setlength{\textfloatsep}{0pt}
\begin{algorithm}[bt]
    \small
	\caption{Fine-grained Localization}
	\label{alg:fine}
	\KwIn{$Q=(d_i,t_q),D_n,L,l_m$} 
	$Stop\_{flag} \leftarrow$ false\;
	$\bar{D}_{n} \leftarrow \varnothing$\;
	\For{$\vDevice[k]\in D_n$}{
		$\bar{D}_{n} \leftarrow d_k$\;
	    \For{$r_j\in R(l_m.loc)$}{
		    Compute $P(r_j|\bar{D}_{n})$;
		}
		\If{$D_n$ \textbf{independent}}{
		Find top-2 probability $P(r_{a}|\bar{D}_{n}),P(r_{b}|\bar{D}_{n})$\; 
		Compute $minP(r_{a}|\bar{D}_{n}),maxP(r_{a}|\bar{D}_{n}),expP(r_{a}|\bar{D}_{n})$\;  
		Compute $minP(r_{b}|\bar{D}_{n}),maxP(r_{b}|\bar{D}_{n}),expP(r_{b}|\bar{D}_{n})$\; 
		\If{$minP(r_{a}|\bar{D}_{n})\geq expP(r_{b}|\bar{D}_{n})$ \textbf{or}  $expP(r_{a}|\bar{D}_{n})\geq maxP(r_{b}|\bar{D}_{n})$}{
		    $Stop\_{flag} \leftarrow true$\;
		}}
		\If{$D_n$ \textbf{dependent}}{
		\If{$\forall \bar{D}_{nl} \subseteq \bar{D}_n$, $\alpha(\{\bar{D}_{nl},d_i\},r_j,t_q)=0$}{
		$Stop\_{flag} \leftarrow true$\;}
		}
		\If{$Stop\_{flag} ==$ true}{break;} 
		\textbf{return} $r_{a}$;
	}
\end{algorithm}

A key question is, \emph{how do we compute these probabilities efficiently?} To compute the maximum probability of $d_i$ being in $r_j$, we can assume that 
all unprocessed devices are in room $r_j$ as described in the theorem below. (See the proofs of theorems in Appendix~\ref{sec:appendix}) .

\vspace{-0.5em}
\begin{mytheory}
\label{theory:max}
	Given a set of already processed devices $\bar{D}_{n}$, a candidate room $r_j$ of $d_i$ ,and the possible world $W$ where all devices $D_{n}\setminus \bar{D}_{n}$ are in room $r_j$, then, $maxP(r_j|\bar{D}_{n}) = Pr(r_j|\bar{D}_{n},W)$.
\end{mytheory}
\vspace{-0.3em}

Likewise, to compute the minimum probability, we can simply assume that none of the unprocessed devices are in room $r_j$. The following theorem states that we can compute the minimum by placing all the unprocessed devices in the room (other than $r_j$) in which $d_i$ has the highest chance of being at time $t_q$. 

\vspace{-0.3em}
\begin{mytheory}
\label{theory:min}
	Given a set of already processed devices $\bar{D}_{n}$, a candidate room $r_j\in R(g_x)$, $r_{max}=argmax_{r_{i}\in R(g_x) \setminus r_j} P(r_{i}|\bar{D}_{n})$, and a possible world $W$ where all devices in $D_{n}\setminus \bar{D}_{n}$ are in room $r_{max}$, then, $minP(r_j|\bar{D}_{n}) = P(r_j|\bar{D}_{n},W)$.
\end{mytheory}
\vspace{-0.3em}
For the expected probability of $r_j$ being the answer of $Q$, we prove that it equals to $P(r_j|\bar{D}_{n})$. 
\vspace{-0.3em}
\begin{mytheory}
\label{theory:expected}
    Given a set of independent devices $D_n$, the set of already processed devices $\bar{D}_{n}$, and the candidate room $r_j$, then,\\ $expP(r_j|\bar{D}_{n}) = P(r_j|\bar{D}_{n})$. 
\end{mytheory}
\vspace{-0.3em}

\noindent\textbf{Relaxing the Independence Assumption.} We next relax the conditional independence assumption we have made so far. In this case, we cannot treat each neighbor device independently. Instead, we divide $\bar{D}_{n}$ into several \textit{clusters} where every neighbor device in a cluster have non-zero group affinity with the rest of the devices. Let $\bar{D}_{nl}\subseteq \bar{D}_{n}$ be a cluster where $\forall d_k,d_{k}^{'} \in \bar{D}_{nl}, \alpha(\{d_k,d_{k}^{'}\},r_j,t_q)>0$. In addition, group affinity of devices of any pair of devices in different clusters equals zero, i.e., $\forall d_k \in \bar{D}_{nl}, d_{k}^{'} \in \bar{D}_{nl^{'}}$, where $l\neq l^{'}$,  $\alpha(\{d_k,d_{k}^{'}\},r_j,t_q)=0$.   
In Figure~\ref{fig:fine-algorithm}(b),  $\bar{D}_{n1}=\{d_2,d_3,d_4\}$ and $\bar{D}_{n2}=\{d_5,d_6\}$. Naturally, we have 
$\bar{D}_{n}=\bigcup_{l}\bar{D}_{nl}$. In this case, we assume that each cluster affects the location prediction of $d_i$ independently. 

Thus, probability
$P(\bar{D}_{n}|r_j)=\prod_{l}P(\bar{D}_{nl}|r_j)$. For each cluster, we compute its conditional probability 
$P(\bar{D}_{nl}|r_j)=\frac{P(\bar{D}_{nl},r_j)}{P(r_j)}$, where $P(\bar{D}_{nl},r_j)=\alpha(\{\bar{D}_{nl},d_i\},r_j,t_q)$. The reason is that $P(\bar{D}_{nl},r_j)$ is the probability that all devices in $\bar{D}_{nl}$ and $d_i$ are in room $r_j$, which equals  $\alpha(\{\bar{D}_{nl},d_i\},r_j,t_q)$ by definition. Thus, 

\vspace{-0.4em}
\begin{equation}
\small
    P(r_j|\bar{D}_{n}) = 1/(1+\frac{1-\prod_{l}\alpha(\{\bar{D}_{nl},d_i\},r_j,t_q)}{1-\alpha(d_i,r_j)})
\end{equation}
\vspace{-0.6em}

\noindent the algorithm terminates when the group affinity for any cluster turns zero.

Finally, we describe the complete fine-grained location cleaning algorithm in Algorithm~\ref{alg:fine}. Given $Q=(d_i,t_q)$, we observe only the neighbor devices at time $t_q$ (Line~4-5). Next, we compute the probability of $P(r_j|\bar{D}_{n})$ for every candidate room in $R(l_m.loc)$ (Line~7-8). If devices are independent, 
we select two rooms with top-2 probability and use loosen stop condition to check if the algorithm converges (Line~10-14). 
Otherwise, we check if all clusters have zero group affinity (Line~15-17). Finally, we output the room when the stop condition is satisfied (Line~13-16). 
\vspace{-0.2em}
\section{LOCATER System}
\label{sec:prototype}
We describe the prototype of LOCATER built based on the previous coarse and fine-grained localization algorithms. Also, we describe a caching engine to scale LOCATER to large connectivity data sets.

\begin{figure}[tb]
    \centering
    \includegraphics[width=1\linewidth]{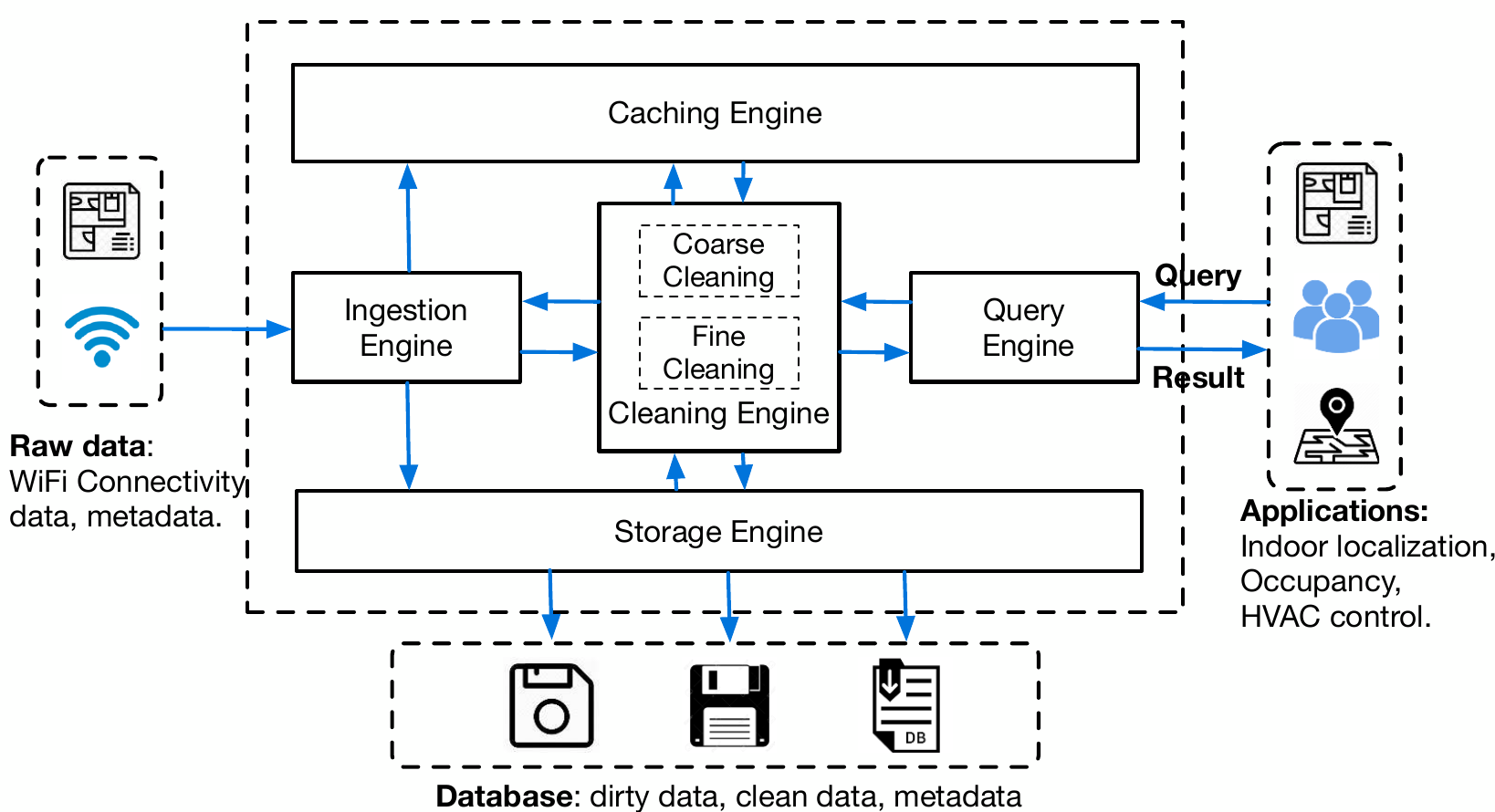}
    \caption{Architecture of LOCATER.}
    \label{fig:architecture}
\end{figure}

\vspace{0.1cm}
\noindent
\textbf{Architecture of LOCATER.} Figure~\ref{fig:architecture} shows the high-level architecture of the LOCATER prototype. LOCATER ingests a real-time stream of WiFi  connectivity events (as discussed in Section~\ref{sec:definition}). 
Additionally, LOCATER takes as input {\em metadata about the space} which includes the {\em set of WiFi APs} deployed in the building, the {\em set of rooms} in the building (including whether each room is a public or private space --see Section~\ref{sec:definition}--), the \textit{coverage of WiFi APs} in terms of list of rooms covered by each AP, and the \textit{temporal validity of connectivity events} per type of device in the building.\footnote{Appendix~\ref{sec:appendix} describes how to obtain this metadata in practice for a real deployment.}

LOCATER supports queries $Q=(d_i,t_q)$ that request the location of device $\vDevice[i]$ at time $t_q$, where $t_q$ could be the current time (e.g., for real-time tracking/personalized services) or a past timestamp (e.g., for historical analysis). 
Given $Q$, LOCATER's {\em cleaning engine} determines if $t_q$ falls in a gap. If so, it executes its coarse-grained localization (Section~\ref{sec:coarse}).
If at $t_q$, $d_i$ was inside the building, the cleaning engine performs the fine-grained localization (Section~\ref{sec:fine}). 
Given a query with associated time $t_q$, LOCATER uses a subset of historical data (e.g., X days prior to $t_q$) to learn  both room and group affinities. We explore the impact of the  amount of historical data used to the accuracy of the model learnt in Section~\ref{sec:evaluation}. 

\vspace{0.1cm}
\noindent
\textbf{Scaling LOCATER.} The cleaning engine computes room and group affinities which requires time-consuming processing of historic data. Algorithm~\ref{alg:fine} iteratively performs such computation for each neighbor device of the queried device. In  deployments with large WiFi infrastructure and number of users, this might involve processing large sets of connectivity events which can be a challenge if applications expect real-time answers.
LOCATER caches computations performed to answer queries and leverages this information to answer subsequent queries. Such cached information constitutes what we will refer to as a \textit{global affinity graph} $\mathcal{G}^{g}=(\mathcal{V}^{g},\mathcal{E}^{g})$, where nodes correspond to devices and edges correspond to pairwise device affinities. 
Given a query $Q=(d_i,t_q)$, LOCATER uses the global affinity graph $\mathcal{G}^{g}$ to determine the appropriate order in which neighbor devices to $d_i$ have to be processed. 
Intuitively, devices with higher device affinity w.r.t. $d_i$ have higher impact on the computation of the fine-grained location of $d_i$ (e.g., a device which is usually collocated with $d_i$ will provide more information about $d_i$'s location than a device than a device that just appeared in the dataset).
We empirically show in our experiments that processing neighbor devices in decreasing order of device affinity instead of a random order makes the cleaning algorithm converge much faster.

\vspace{0.05in}
\noindent\emph{(1) Building the local affinity graph.} 
The affinities computed in Section~\ref{sec:fine} can be viewed as a graph, which we refer to as {\em local affinity graph} $\mathcal{G}^{l}=(\mathcal{V}^{l},\mathcal{E}^{l})$, where $\mathcal{V}^{l}=\bar{D}_n \cup d_i$.
In this time-dependent local affinity graph, each device in $\bar{D}_n$, as well as the queried device $d_i$, are nodes and the edges represent their affinity. Let $e^{l}_{ab} \in \mathcal{E}^{l}$ be an edge between nodes $d_a$ and $d_b$ and $w(e^{l}_{ab},t_q)$ be its weight measuring the probability that $d_a$ and $d_b$ are in the same room at time $t_q$. The value of $w(e^{l}_{ab},t_q)$ is computed based on Algorithm~\ref{alg:fine} as $w(e^{l}_{ab},t_q)=\frac{\sum_{r_{j}\in R(g_x)}\alpha(\{d_a,d_b\},r_j,t_q)}{|R(g_x)|}$. 

\begin{figure}[bt]
	\centering
	\includegraphics[width=0.65\linewidth]{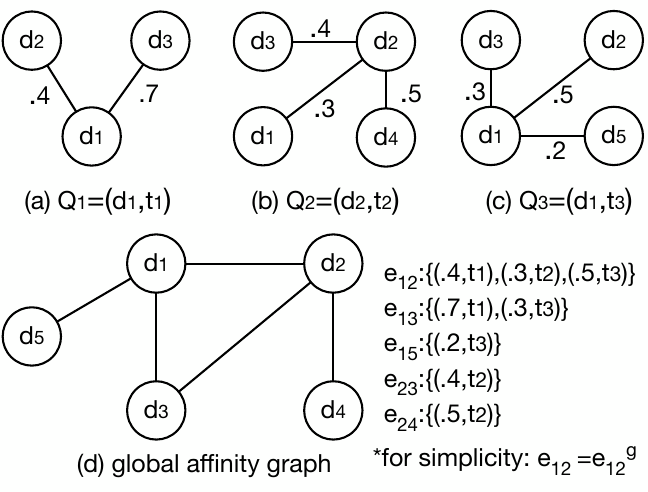}
	\caption{Generation of global affinity graph (d) from local affinity graphs (a,b,c).}
	\label{fig:cache}
\end{figure}

\vspace{0.05in}
\noindent\emph{(2) Building the global affinity graph.} 
After generating a local affinity graph for $d_i$ at time $t_q$, this information is used to update the global affinity graph. We will illustrate the process using Figure~\ref{fig:cache}. Given the current global affinity graph $\mathcal{G}^{g}=(\mathcal{V}^{g},\mathcal{E}^{g})$ and a local affinity graph $\mathcal{G}^{l}=(\mathcal{V}^{l},\mathcal{E}^{l})$, the updated global affinity graph $\mathcal{G}^{g}=(\hat{\mathcal{V}^{g}},\hat{\mathcal{E}^{g}})$ is such that $\hat{\mathcal{V}^{g}}=\mathcal{V}^{g}\cup \mathcal{V}^{l}$ and $\hat{\mathcal{E}^{g}}=\mathcal{E}^{g}\cup \mathcal{E}^{l}$. Note that, as affinity graphs are time-dependent, in the global affinity graph we associate each edge included from an affinity graph with its timestamp $t_q$ along with its weight. Hence, in the global affinity graph, the edge in between two nodes is a vector which stores the weight-timestamp pairs associated with different local affinity graphs $v^{g}_{ab}=\{(w(e^{l}_{ab}),t_1),...,(w(e^{l}_{ab}),t_n)\}$. When merging the edge set, we merge corresponding vectors -- i.e., $v^{g}_{ab}=v^{g}_{ab}\cup w(e^{l}_{ab},t_q)$ for every $e^{g}_{ab} \in \mathcal{E}^{g}$. For instance, in the global affinity graph in Figure~\ref{fig:cache}(d), which has been constructed from three different local affinity graphs (Figure~\ref{fig:cache}(a),(b),(c)), the edge that connects nodes $d_1$ and $d_2$ has the weight-timestamp values extracted from each local affinity graph $(.4,t_1), (.3,t_2),(.5,t_3)$. To control the size of the global affinity graph, we could delete  past affinities stored in the graph $(w(e^{l}_{ab}),t_i)$ , $\tau - t_i > T_s$, where $\tau$ is current time and $T_s$ is a threshold defined by users, e.g., 3 months.

\vspace{0.05in}
\noindent\emph{(3) Using the global affinity graph.}  
When a new query $Q=(d_i,t_q)$ is posed, our goal is to identify the neighbor devices that share high affinities with $d_i$ and use them to compute the location of $d_i$ using Algorithm~\ref{alg:fine}. Given the set $D_n$ of  devices that are neighbors to  $d_i$ at time $t_q$, we compute the affinity between $d_i$ and each device $d_k \in D_n$, denoted by $w(e^{g}_{ik},t_q)$, using the global affinity graph. As each edge in the global affinity graph contains a vector of affinities with respect to time, we compute affinity by assigning a higher value to those instances that are closer to the query time $t_q$ as follows: $w(e^{g}_{ik},t_q)=\sum_{j=1}^{j=n}l_{j}w(e^{l}_{ik},t_j)$, where $l_j$ follows a normal distribution, $\mu=t_q$ and $\sigma^{2}=1$ that is normalized. Finally, we create a new set of neighbor devices $\mathcal{N}^{g}(d_i)$ and include each device $d_k \in D_n$ in descending order of the computed affinity $w(e^{g}_{ik},t_q)$. This new set replaces $D_n$ in Algorithm~\ref{alg:fine}. Thus, the algorithm processes devices in descending order of affinity in the global affinity graph.

\vspace{-1em}
\section{Evaluation}
\label{sec:evaluation}

\vspace{-0.2em}
We implemented a prototype of LOCATER and performed experiments to test its performance in terms of quality of the cleaned data, efficiency, and scalability. The experiments were executed in an 8 GB, 2 GHz Quad-Core Intel Core i7 machine with a real dataset as well as a synthetic one. We refer to the implementation of LOCATER's fine-grained algorithms based on independent and relaxed independent (dependent) assumptions as I-FINE and D-FINE. Correspondingly, we will refer to the system using those algorithms as I-LOCATER and D-LOCATER, respectively. 

\vspace{-0.5em}
\subsection{Experimental Setup}
\vspace{-0.2em}
\noindent\textbf{Dataset.} We use connectivity data captured by the TIPPERS system~\cite{mehrotra2016tippers} in our DBH building at UC Irvine, with 64 WiFi APs, 300+ rooms (including classrooms, offices, conference rooms, etc.) and an average daily occupancy of about 3,000. On average, each WiFi AP covers 11 rooms. The dataset (in the following \textsf{DBH-WIFI}) contains 10 months of data, from Sep. 3rd, 2018 to July 8th, comprising $38,670,714$ connectivity events for $66,717$ different devices.

\noindent
\textbf{Ground truth.} 
We collect fine-grained location of 28 distinct individuals as ground truth. We asked 9 participants to log their daily activity within the building (the room where they were located and how much time they spent in it) for a week. The participants filled in comprehensive and precise logs of their activity amounting to 422 hours in total. We also selected three cameras in the building that cover different types of spaces (i.e., faculty offices area, student offices area, and lounge space). We manually reviewed the camera footage to identify individuals in it (the area covered is in our portion of the building so we identified 26 individuals -- 7 of them were also participants of the daily activity logging--) and their locations. We requested the identified individuals for their MAC address. If a person $p$ with MAC address $m$ was observed to enter a room $r$ at time $t_1$ and leaving the room at time $t_2$, we created an entry in our ground truth locating $m$ in room $r$ during the interval $(t_1,t_2)$.

\noindent
\textbf{Queries.} We generated a set of $10,028$ queries, denoted by $\mathcal{Q}$, related to individuals in the ground truth ($3,129$ queries for participants that logged their activities and $6,899$ queries for individuals detected in the camera images). The number of queries per individual are approximately the same, as far as differences in the labeled elements per user allow it.

\noindent
\textbf{Baselines.} Traditional indoors localization algorithms are either based on active localization or passive localization using information such as signal strength maps (as explained in Section~\ref{sec:intro}). Hence, we defined two baselines used in practice for the kind of semantic localization described in this paper (i.e., coarse and fine-grained localization based on connectivity logs and background information). The baselines are defined as follows: \emph{Baseline1} and \emph{Baseline2} use \emph{Coarse-Baseline} for coarse localization and for fine-grained localization they use \emph{Fine-Baseline1} and \emph{Fine-Baseline2},  respectively. In \emph{Coarse-Baseline}, the device is considered outside if the duration of a gap is at least one hour, otherwise the device is inside and the predicted region is the same as the last known region. \emph{Fine-Baseline1} selects the predicted room randomly from the set of candidates in the region whereas \emph{Fine-Baseline2} selects the room associated to the user based on metadata (e.g., his/her office).

\noindent
\textbf{Quality metric.} 
LOCATER can be viewed as a multi-class classifier whose classes correspond to all the rooms and a label for outside the building. We use the commonly used {\em accuracy} metric~\cite{tharwat2020classification}, defined next, as the measure of quality.\footnote{Accuracy, as defined in the paper, is exactly the same as other micro-metrics such as micro-precision, recall, and F-measure~\cite{sokolova2009systematic}. Micro-level metrics are, often, more reflective of overall quality of the multi-level classifier (such as LOCATER) when the query dataset used for testing is biased towards some classes.} Let $\mathcal{Q}$ be the set of queries,  $\mathcal{Q}_{out}, \mathcal{Q}_{region}, \mathcal{Q}_{room}$ be the subset of queries for which LOCATER returns correctly the device's location as being \emph{outside}, in a specific region, and a specific room, respectively. 
Accuracy of the coarse-grained algorithm can then be measured as: $A_{c}=(|\mathcal{Q}_{out}|+|\mathcal{Q}_{region}|)/|\mathcal{Q}|$. Likewise, for fine-grained and overall algorithm, accuracy corresponds to   $A_{f}=|\mathcal{Q}_{room}|/|\mathcal{Q}_{region}|$, and  $A_{o}=(|\mathcal{Q}_{room}|+|\mathcal{Q}_{out}|)/|\mathcal{Q}|$, respectively.

\begin{figure}[bt]
	\centering
	\begin{minipage}[t]{0.49\linewidth}
	\vspace{-0.4em}
		\subfigure{\includegraphics[width=4.1cm,height=3.4cm]{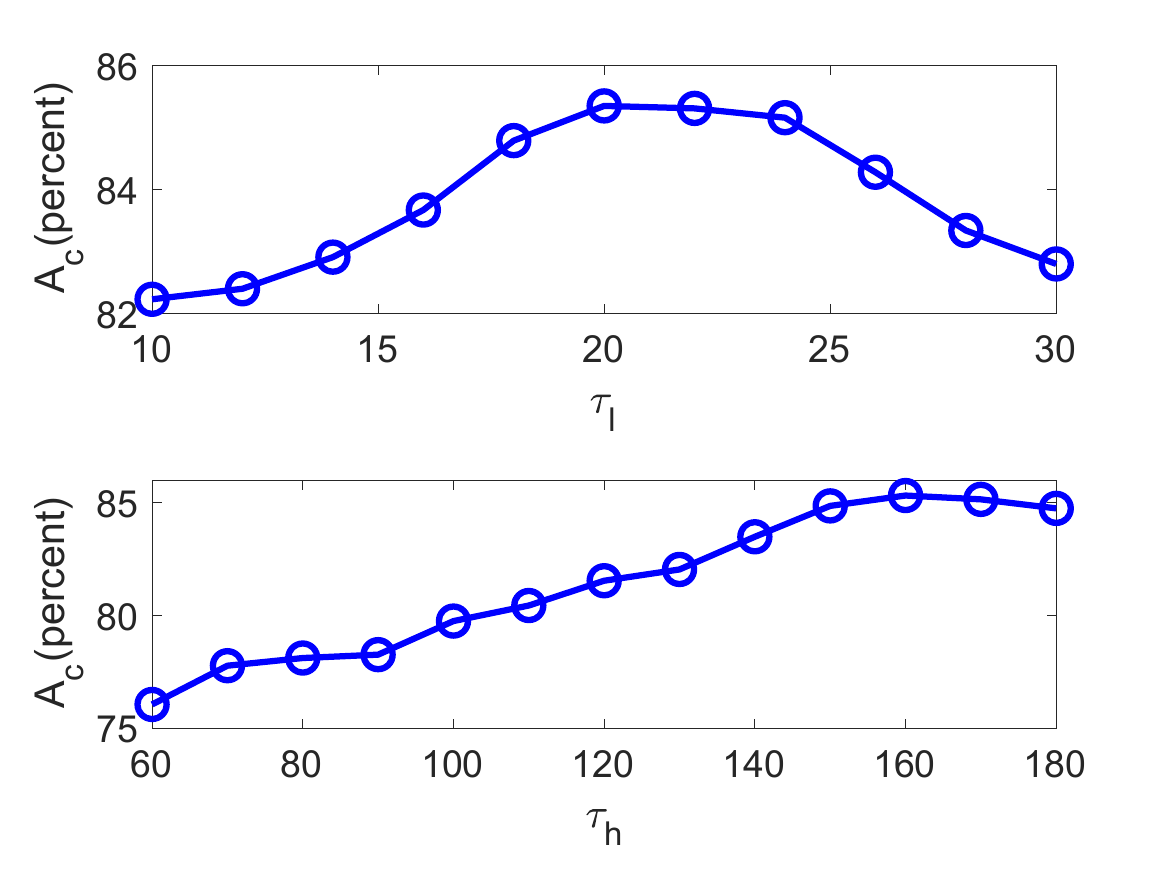}}
		\vspace{-0.6em}
	\caption{\small Thresholds tuning.}
	\label{fig:coarse_config}
	\end{minipage}
	\begin{minipage}[t]{0.49\linewidth}
		\subfigure{\includegraphics[width=4.1cm,height=3.2cm]{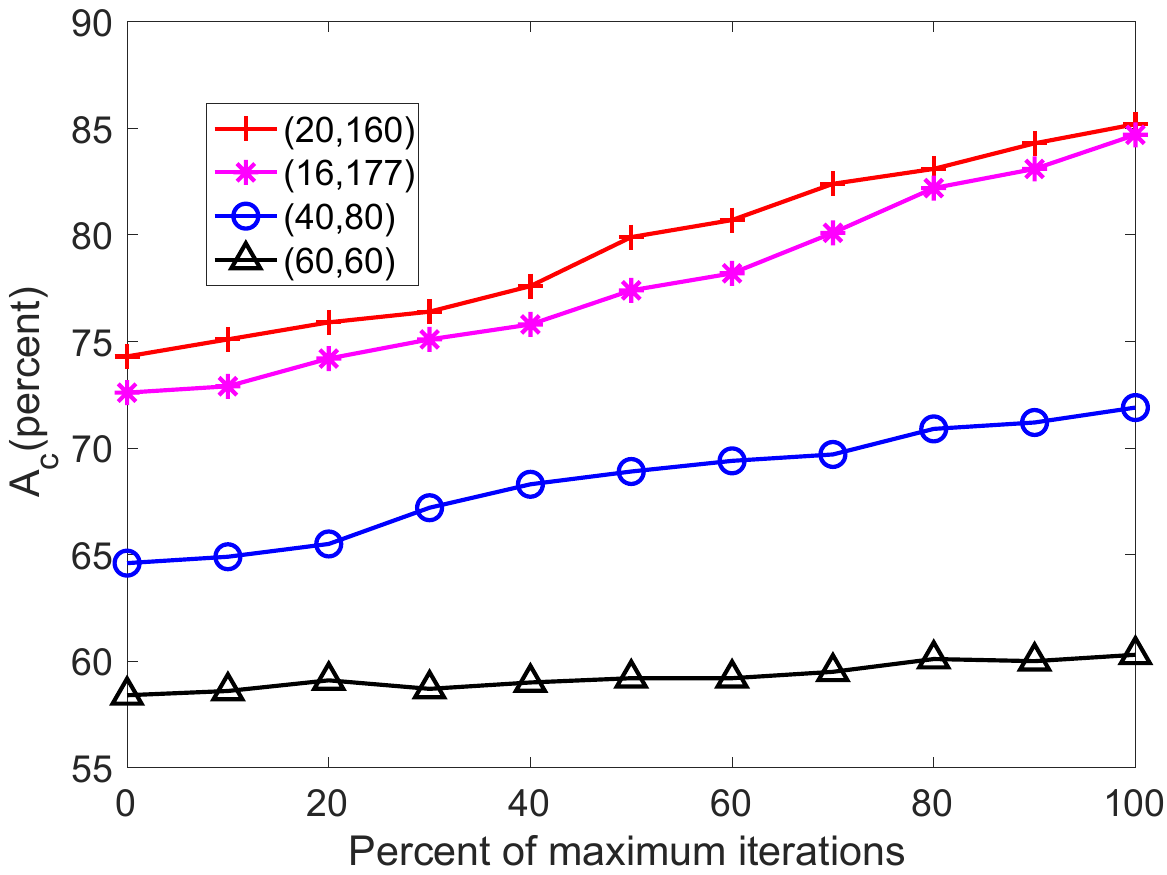}}
		\vspace{-0.3em}
	\caption{\small Iteration.}
	\label{fig:iteration}
	\end{minipage}
		\vspace{-0.2em}
\end{figure}

\begin{figure*}[htbp]
	\centering
	\begin{minipage}[t]{0.74\linewidth}
	\vspace{-0.5em}
		\subfigure
		{\label{fig:coarse_robust}\includegraphics[width=4.2cm,height=3.3cm]{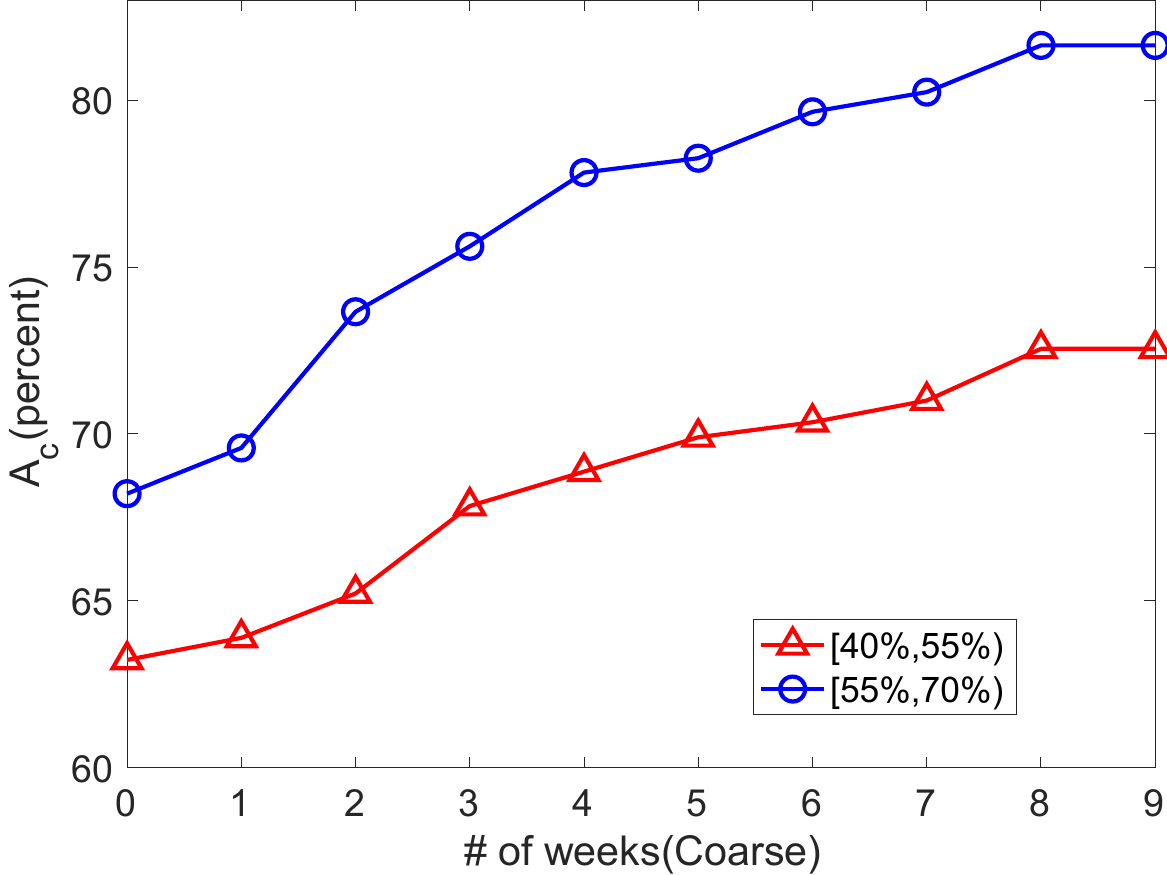}}
		\subfigure
		{\label{fig:fine_robust}\includegraphics[width=4.2cm,height=3.3cm]{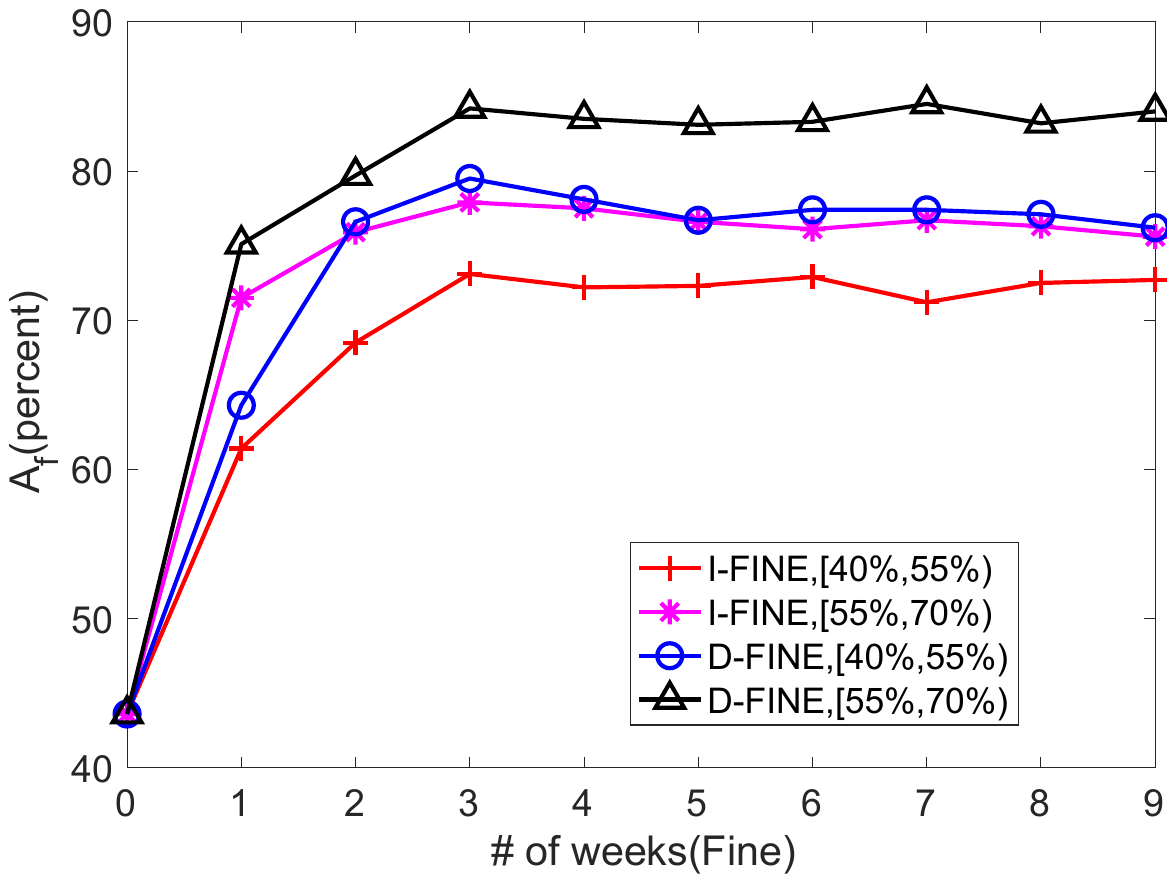}}
		\subfigure
		{\label{fig:overall_robust}\includegraphics[width=4.2cm,height=3.3cm]{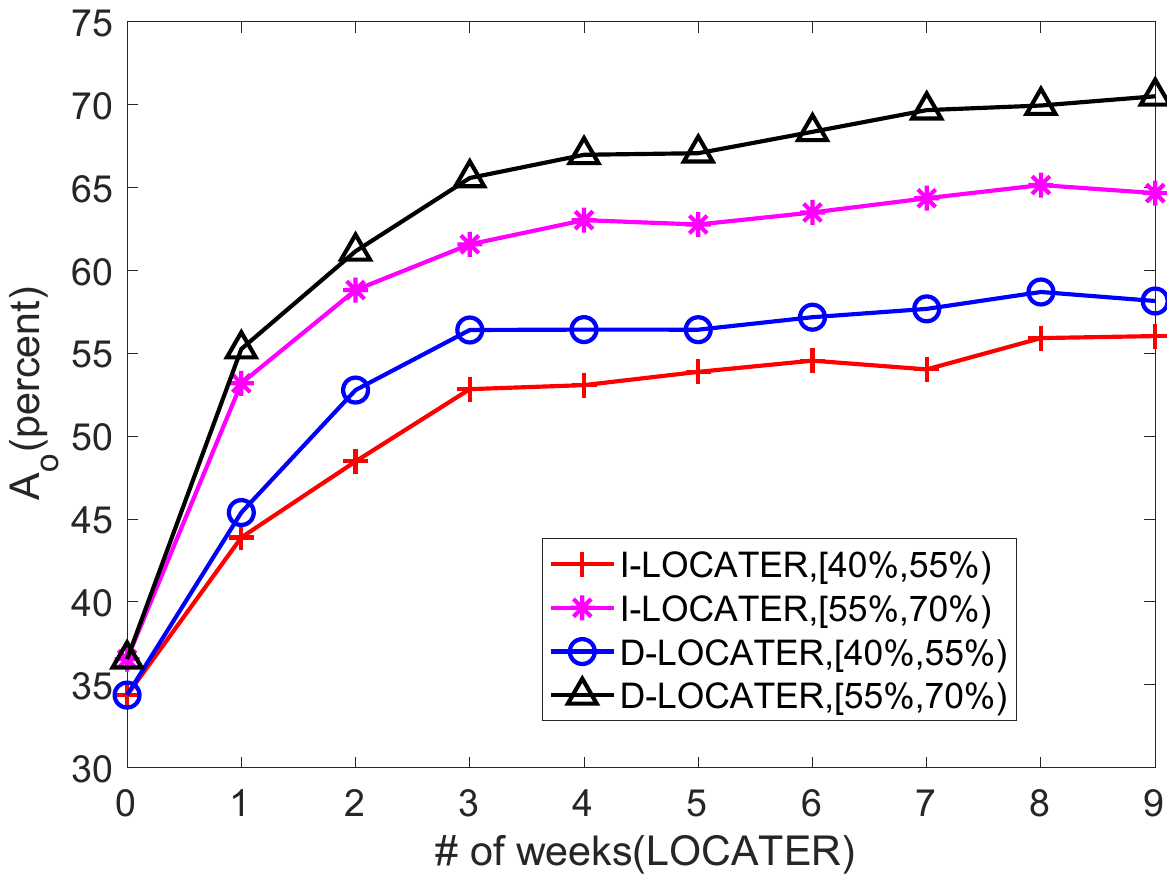}}
		\vspace{-0.5em}
		\caption{\small Impact of historical data used on accuracy.}
		\label{fig:robust}
		\vspace{-0.5em}
	\end{minipage}
	\begin{minipage}[t]{0.25\linewidth}
		\subfigure	{\includegraphics[width=4.2cm,height=3.3cm]{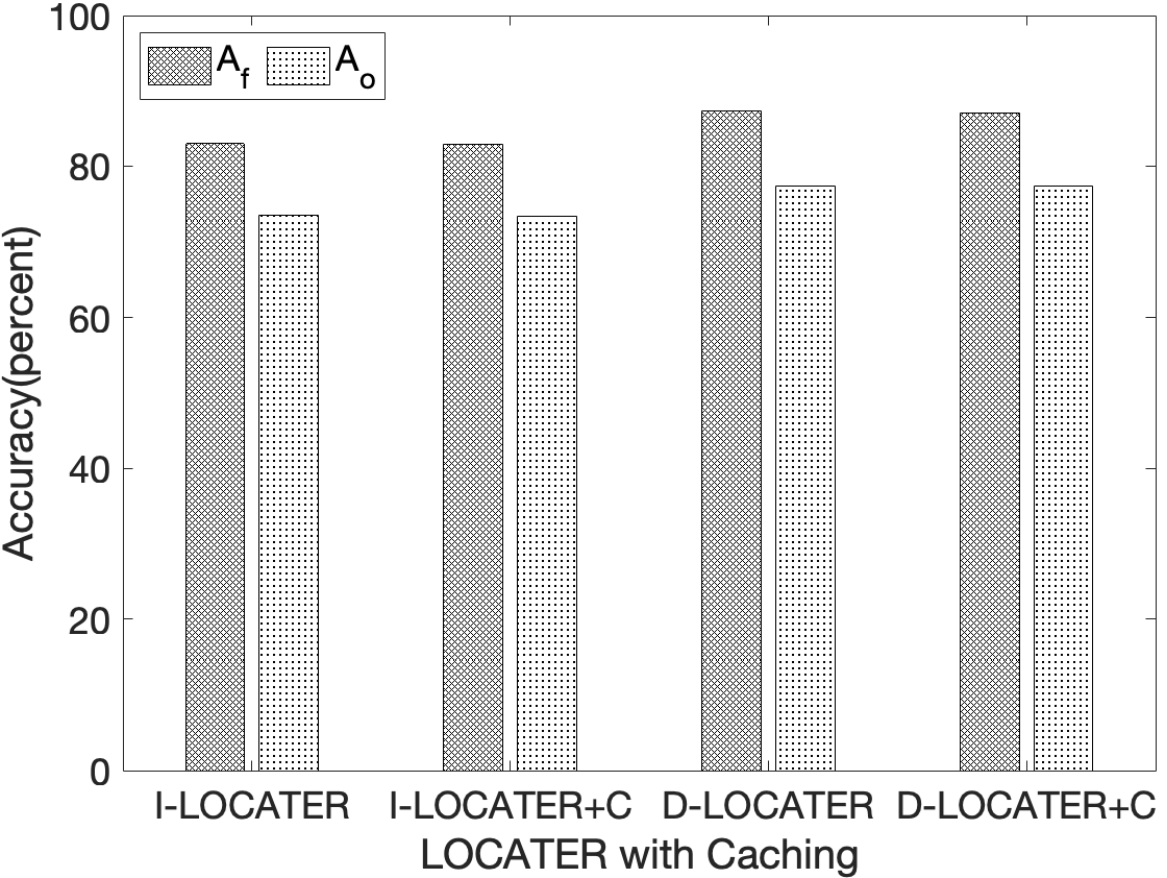}}
	\vspace{-0.5em}
	\caption{\small Caching accuracy.}
	\label{fig:compare-caching}
		\vspace{-0em}
	\end{minipage}
	\vspace{-0.5em}
\end{figure*}

\vspace{-1em}
\subsection{Accuracy on \textsf{DBH-WIFI} Dataset}
\label{sec:testRealData}
\vspace{-0.5em}

We first test the performance of LOCATER, in terms of accuracy, for the \textsf{DBH-WIFI} dataset. As LOCATER exploits the notion of recurring patterns of movement/usage of the space, we analyze the performance w.r.t. the level of \emph{predictability} of different user profiles. We consider the fact that some people spend most of their time in the building in the same room (e.g., their offices) as a sign of predictable behaviour. We can consider this as their ``preferred room". We group individuals in the dataset into 4 classes based on the percentage of time they spend in their preferred room: $[40,55)$, $[55,70)$, $[70,85)$ and $[85,100)$, where $[40,55)$ means that the user spent 40-55 percent of time in that room (no user in the ground truth data spent less than $40\%$ of his/her time in a specific room). 

\noindent\textbf{Impact of thresholds in coarse localization.} 
The coarse-level localization algorithm depends upon two thresholds: $\vLowBuild$ and $\vHighBuild$. We use $k$-fold cross validation with $k=10$ to tune them. We
vary $\vLowBuild$'s value from 10 to 30 minutes and $\vHighBuild$'s value from 60 to 180 minutes. We fix $\vHighBuild=180$ when running experiments for $\vLowBuild$ and fix $\vLowBuild=20$ when running experiments for $\vHighBuild$. 
From Figure~\ref{fig:coarse_config} we observe that, with the increasing of $\vLowBuild$, the accuracy increases first and then slightly decreases after it peaks at $\vLowBuild=20$. For $\vHighBuild$, when it increases, accuracy gradually increases and levels off when $\vHighBuild$ is beyond 170. 
We also test the parameters computed by confidence interval in Section~\ref{subsec:parameter-computation}, which are $\vLowBuild=16.4$ and $\vHighBuild=177.3$. The accuracy achieved by this parameter setting is $84.7\%$, which is close to the best accuracy ($85.2\%$) achieved by parameters tuned based on cross validation. 

\noindent\textbf{Iterative classification for coarse localization}
We test the robustness of the iterative classification method. We vary the quality of the initial decisions of the heuristic strategy (without iterations) by  setting the parameters $(\vLowBuild,\vHighBuild)$ to $(20,160)$, $(16,177)$,
$(40,80)$, and $(60,60)$. For each query we terminate the coarse localization algorithm at different stages (as a percentage of the maximum iterations the algorithm would perform) and report $A_c$ in Figure~\ref{fig:iteration}. We observe that for a high quality initial decision, the iterative classification improves the accuracy significantly with increasing number of iterations. Also, for those relatively bad initial decisions (with initial accuracy $58\%$ and $65\%$) the improvement achieved by the iterative classification is small but it always increases. We also show that for the parameters decided by the Gaussian confidence interval method (i.e., $(16,177)$), which does not rely on the ground truth data, the iterative classification method works very well. 

\noindent\textbf{Impact of weights of room affinity.}
We examine the impact of weights used in computing room affinity, $w^{pf},w^{pb},w^{pr}$.
We report the fine accuracy of the four weight combinations satisfying the rules defined in that section: $C1=\{0.7,0.2,0.1\}$, $C2=\{0.6,0.3,0.1\}$, $C3=\{0.5,0.3,0.2\}$, and $C4=\{0.5,0.4,0.1\}$. 
For $C_1, C_2, C_3, C_4$, $A_f$ of I-FINE is 81.8, 83.4, 82.3, 82.4, and $A_f$ of D-FINE is 86.1, 87.5, 86.6 and 86.4, respectively. 
We observe that all the combinations for both I-FINE and D-FINE obtain a similar accuracy (with $C2$ achieving a slightly higher accuracy). Hence, the algorithm is not too sensitive to the weight distributions in this test. Also, D-FINE outperforms I-FINE by 4.6\% on average.

\noindent\textbf{Impact of historical data.} 
We use historical data to train the models in the coarse algorithm and to learn the affinities in the fine algorithm.   
We explored how the amount of historical data used affects the performance of LOCATER. We report the coarse, fine, and overall accuracy for the [40,55)$\%$ and [55,70)$\%$ predictability groups, in Figure~\ref{fig:coarse_robust}, Figure~\ref{fig:fine_robust}, and Figure~\ref{fig:overall_robust},  respectively. The graphs plot the accuracy of the algorithm with increasing amount of historical data, from no data at all up to 9 weeks of data. The accuracy of the coarse-grained algorithm increases with increasing amount of historical data and it reaches a plateau when 8 weeks of data are used. The reason is that the iterative classification algorithm used to train the model becomes more generalized the more data is used for the training. 
The performance of the fine-grained algorithm is poor when no historical data is used (as this effectively means selecting the room just based on its type). However, when just one week of historical data is used the performance almost doubles. The accuracy keeps increasing with increasing number of weeks of data though the plateau is reached at 3 weeks. The results show that the kind of affinities computed by the algorithm are temporally localized. The overall performance of the system follows a similar pattern. With no data, mistakes made by the fine-grained localization algorithm penalize the overall performance. With increasing amount of historical data, the performance increases due to the coarse-grained algorithm labeling gaps more correctly. In all the graphs, the performance of the overall system and its algorithms increases with increasing level of predictability of users.

\vspace{0.1cm}

\noindent{\textbf{Robustness of LOCATER w.r.t. room affinity.}
LOCATER's approach to disambiguating locations  exploits prior probability of individuals to be in  specific rooms (room affinity). In this experiment, we explore the robustness of LOCATER when we only know the prior for a smaller percentage of  people. 
We randomly select users for whom we compute and associate a room affinity to each candidate room (based on historical data and room metadata). For the rest, we consider an uniform room affinity for all the candidate rooms. We repeat the experiment 5 times and report the average fine accuracy: $A_f$. We set the percentage of users with refined room affinities to 0\%, 25\%, 50\%, 75\%, and 100\%, and the corresponding $A_f$ is: $6.2$, $57.1$, $71.3$, $81.1$, $87.1$. We observe that the accuracy is poor when equally distributed affinity is considered for all users. When a refined room affinity is computed for  a small portion of users (25\%), the accuracy increases significantly to $57.1$.
Increasing the number of users with refined room affinity makes the accuracy converge to $87.1$. 
Thus, we expect LOCATER to work very well in scenarios where the pattern of building usage and priors for a significant portion of the occupants is predictable.

\noindent\textbf{Impact of caching.} We examine how the fine-grained algorithm's caching technique (see Section~\ref{sec:fine}) affects the accuracy of the system. We compute the accuracy of both I-LOCATER and D-LOCATER compared to their counterparts using caching I-LOCATER+C and D-LOCATER+C. Figure~\ref{fig:compare-caching} plots the overall accuracy of the system averaged for all the tested users. We observe that adding caching incurs in a reduction of the accuracy from 5\%-10\%, which does not significantly affect the performance. This means that the device processing order generated by the caching technique maintains a good accuracy while decreasing the cleaning time (see Section~\ref{sec:testEfficiency}). 

\noindent \textbf{Probability distribution of results.} We show the probability distribution computed by LOCATER for each of the rooms in the set of candidate rooms for a given query. In particular, we plot the highest probability value associated with any room ($Pr_h$), the difference of the highest and second highest probability ($\Delta Pr$), and the summation of the remaining probabilities ($\sum_{r}$). We report the statistics over all the queries in Table~\ref{tab:probabilityDistribution}. We observe a long tail distribution for the set of different rooms output by LOCATER. In particular, there are $69\%$ queries whose highest probability is in $[.4,.6)$, $43\%$ queries whose difference of the highest and second highest probability is $[.2,.3)$ and $51\%$ queries where the sum of top-2 probabilities is greater than $.6$. 


\begin{table}[bt]
	\small
	\centering
	\caption{Probability distribution of rooms.}
	\vspace{-0.1em}
	\label{tab:probabilityDistribution}
	\setlength\tabcolsep{3.5pt}
	\begin{tabular}{|c|c|c|c|c|c|}
		\hline
		  $Pr_{h}$ & $[0,.2)$ & $[.2,.4)$ & $\textbf{[0.4,.6)}$ & $[.6,.8)$ & $[.8,1)$\\ \hline
		 Percent of queries & $0$ & $19$ & $\textbf{69}$ & $12$ & $0$ \\ \hline
		 $\Delta Pr$ & $[0,.1)$ & $[.1,.2)$ & $\textbf{[.2,.3)}$ & $[.3,.4)$ & $[.4,.5)$ \\ \hline 
		 Percent of queries & $4$ & $17$ & $\textbf{43}$ & $20$ & $16$ \\ \hline
		 $\sum_{r}$ & $[0,.2)$ & $\textbf{[.2,.4)}$ & $[.4,.6)$ & $[.6,.8)$ & $[.8,1)$ \\ \hline 
		  Percent of queries & $32$ & $\textbf{51}$ & $15$ & $2$ & $0$ \\ \hline
	\end{tabular}
\end{table}

\noindent\textbf{Comparison with baselines.} We compare accuracy of LOCATER vs. baselines for different predictability groups and overall (as the average of accuracy for all people) as $\mathcal{Q}$ (see Table~\ref{tab:compareBaseReal} where each cell shows the rounded up values for $A_c,A_f,A_o$).
We observe that both I-LOCATER and D-LOCATER significantly outperform \emph{Baseline1} regardless of the predictability level of people. This is due to the criteria to select the room in which the user is located when performing fine-grained localization. Deciding this at random works sometimes in situations where the AP covers a small set of large rooms but incurs in errors in situations where an AP covers a large set of rooms (e.g., in our dataset up to 11 rooms are covered by the same AP). \emph{Baseline2} uses an strategy where this decision is made based on selecting the space where the user spends most of his/her time, if that space is in the region where the user has been localized. This strategy only works well with very predictable people. Hence, LOCATER outperforms \emph{Baseline2} in every situation except for the highest predictable group where \emph{Baseline2} obtains a slightly better accuracy. The accuracy of D-LOCATER is consistently higher than I-LOCATER. Both of them perform significantly better than the baselines except for the situation highlighted before.


\begin{table}[bt]
	\small
	\centering
	\caption{Accuracy for different predictability groups.}
	\label{tab:compareBaseReal}
	\setlength\tabcolsep{2.5pt}
	\begin{tabular}{|c|c|c|c|c|c|}
		\hline
		$A_c|A_f|A_o$ & $[40,55)$ & $[55,70)$ & $[70,85)$ & $[85,100)$ & $\mathcal{Q}$ \\ \hline
		\emph{Baseline1} & $56|10|24$ & $63|8.0|25$ & $67|10|26$ & $73|12|28$ & $\textbf{64}|\textbf{10}|\textbf{26}$ \\ \hline
		\emph{Baseline2} & $62|45|39$ & $67|63|50$ & $69|75|57$ & $76|93|72$ & $\textbf{68}|\textbf{67}|\textbf{53}$ \\ \hline
		I-LOCATER & $76|72|61$ & $83|78|70$ & $87|84|77$ & $93|87|84$ & $\textbf{85}|\textbf{83}|\textbf{75}$ \\ \hline
		D-LOCATER & $76|77|63$ & $83|82|72$ & $87|87|79$ & $93|92|88$ & $\textbf{85}|\textbf{87}|\textbf{79}$  \\ \hline
	\end{tabular}
\end{table}

\begin{table}[bt]
	\small
	\centering
	\vspace{-0.8em}
	\caption{Macro results of LOCATER for different methods.}
	\label{tab:macro}
	\setlength\tabcolsep{3.5pt}
	\begin{tabular}{|c|c|c|c|c|} 
	\hline
	 & Precision & Recall & F-1 \\ \hline 
	Baseline1 & $21.8$ & $33.5$ & $26.4$  \\ \hline
	Baseline2 & $58.7$ & $46.2$ & $51.7$  \\ \hline
	I-LOCATER & $78.2$ & $73.7$ & $76.7$  \\ \hline
	D-LOCATER & $81.3$ & $76.4$ & $78.8$  \\ \hline
	\end{tabular}
\end{table}

\noindent\textbf{Macro results.}
 We report macro precision, recall, and F-1 measure for \emph{Baseline1}, \emph{Baseline2}, I-LOCATER, and D-LOCATER, respectively. Macro precision (recall) is defined as the average of precision (recall) of all classes. As shown in Table~\ref{tab:macro}, LOCATER achieved a significantly better precision and recall than baselines and the performance of D-LOCATER is slightly better than I-LOCATER's.

\vspace{-0.5em}
\subsection{Efficiency and Scalability}
\label{sec:testEfficiency}

We first examine the efficiency of LOCATER on the \textsf{DBH-WIFI} dataset. We report average time per query when the system uses or not the stopping conditions described in Section~\ref{sec:fine}. With stop condition, LOCATER takes 563ms while it takes 2,103ms without it.
Without stop conditions, I-LOCATER has to process all neighbor devices, whereas with the stop conditions the early stop brings a considerable improvement in the execution time. 

We conduct scalability experiments both on real and synthetic data. We randomly select a \textit{subspace} of a building by controlling its size using as parameters the number of WiFi APs, rooms, and devices. For the real dataset, \textsf{DBH-WIFI}, we extract four datasets, $Real_1$, ..., $Real_4$. The number of WiFi APs for these four datasets are 10, 30, 50,  64, and the number of rooms are 46, 152, 253 and 303, and the number of devices are 41,343, 60,885, 63,343, 64,717, respectively. 
To test the scalability of LOCATER on various scenarios, we generated four synthetic datasets simulating the following environments, which we list in order of increasing predictability: airport, mall, university, and office.
For each of them we used a real blueprint (e.g., Santa Ana's airport for the first scenario) and created types of people (e.g., TSA staff, passengers, etc) and events they attend (e.g., security checks, boarding flights, etc.) based on our observations. 
Due to space limitation, we only 
report the running time of LOCATER on \textsf{Mall} scenario. In particular, we generated four synthetic datasets, $Mall_1$, ..., $Mall_4$.\footnote{The synthetic data sets have also been used to evaluate the generality of LOCATER to different settings. We showed detailed information about the specific simulated scenarios (including how were they generated) and the complete results (including accuracy for baselines and LOCATER) in the extended version of the paper~\cite{locater2020}.}

We plot the average time cost per query on \textsf{DBH-WIFI} and \textsf{Mall} in Fig~\ref{fig:scale}. The main observations from the results on both datasets are: 1) The caching strategy decreases the computation time of D-LOCATER significantly, and D-LOCATER performs slightly bettern than I-LOCATER; 2) With the caching technique LOCATER has a great scalability when the size of space increases to large scale to support a near-real time query response. (Around 1 second for D-LOCATER and half a second for I-LOCATER).

\begin{figure}
\subfigure{\includegraphics[width=4.1cm,height=3.2cm]{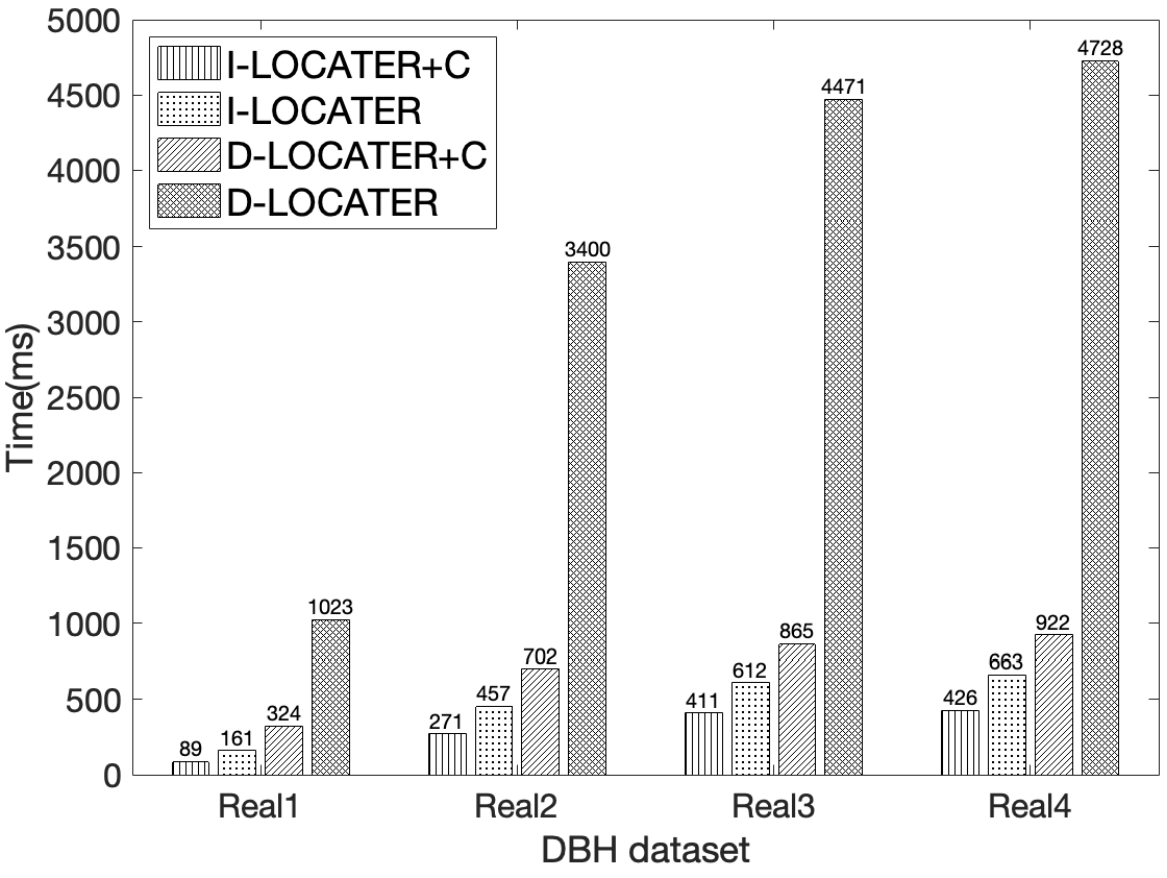}}
	\subfigure{\includegraphics[width=4.1cm,height=3.2cm]{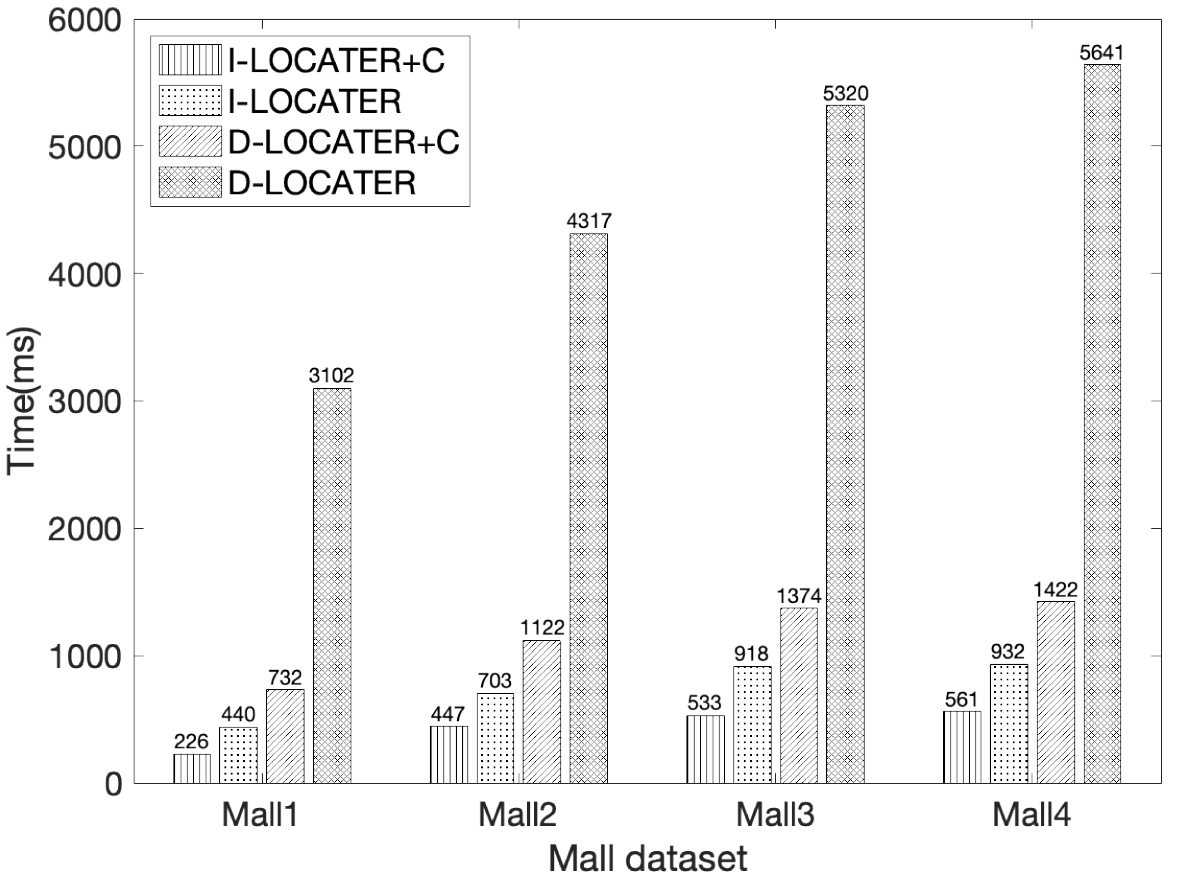}}
	\vspace{-0.5em}
\caption{\small Scalability testing} 
\label{fig:scale}
\end{figure}


\vspace{-0.5em}
\section{Related Work}
\label{sec:related}

LOCATER's semantic localization technique is  related to prior data cleaning work on  missing value imputation, imputing missing time series data ~\cite{khayati2020mind,wellenzohn2017continuous,yi2016st,li2009dynammo, yu2016temporal,mei2017nonnegative,balzano2018streaming,khayati2014memory}, and reference disambiguation. 
Broadly, missing value repair can be classified as rule-based~\cite{fan2012towards,song2015enriching,song2018enriching} - that fills missing values based on the corresponding values in neighboring tuples based on rules; external source-based~\cite{fan2012towards,yakout2011guided,chu2015katara,shan2019webput} - that exploits external data sources such as knowledge bases; and  statistics-based~\cite{mayfield2010eracer,yakout2013don,de2016bayeswipe} - that exploits statistical correlations amongst attributes to repair data. External data source and rule based techniques are unsuitable in our setting since we would like our method to work with minimal assumptions about the space and its usage. For a similar reason, existing statistical approaches, which learn a model based on part of the data known to be clean (using a variety of ML techniques such as relational dependencies) and use it to iterate and fill in missing values, do not apply to the setting of our problem. We do not have access to clean data and, moreover, our approach is based on exploiting temporal features in the data to predict a person's missing location.


Prior work on reference disambiguation~\cite{kalashnikov2005exploiting,dong2005reference,bhattacharya2007collective,li2010multiple} has explored resolving ambiguous references to real-world entities in tuples exploiting tuple context, external knowledge, and relationships implicit in data. 
If we consider the region in the location field in our context to be a reference to a room in the region, fine-grained location disambiguation can be viewed as an instance of reference disambiguation.
Of the prior work, ~\cite{kalashnikov2005exploiting}, that exploits strength of relationships between entities for disambiguation, is the most related to our approach. 
In contrast to~\cite{kalashnikov2005exploiting,dong2005reference,bhattacharya2007collective,li2010multiple}, that focus on cleaning a complete static collection of data, we clean only the tuples needed to answer the location query for an individual. Cleaning the entire semantic location table will be prohibitively expensive since sensor data arrives continuously. Also, our approach exploits two specific relationships --  people's affinity to rooms, and possibility of people to be collocated over time -- that can be relatively easily determined from building metadata and lifted from prior sensor data.
Our algorithm is based on a probabilistic model which also differs from prior work that has taken a more heuristic approach to measuring relationship strengths. Finally, in our setting, temporal properties of data (such as recency) play an important role for disambiguation which has not been considered in prior work on exploiting relationships for disambiguation -- e.g.,~\cite{kalashnikov2005exploiting}. 

Cleaning of sensor data has previously been studied in the context of  applications such as   object-tracking  ~\cite{chu2016data,8943205,zhang2017time,deligiannakis2009another,xu2018improved,baba2016learning,subramaniam2006online,jeffery2006pipelined} that have considered  statistical methods to detect and repair cross readings and missing readings in RFID signals ~\cite{jeffery2006adaptive,xu2018improved,baba2016learning} and techniques to detect outliers in sensor readings~\cite{deligiannakis2009another,subramaniam2006online}. These techniques are specific to RFID data and, as such, do not apply to cleaning WiFi connectivity data.

Indoor localization techniques are broadly based on (a) exploiting (one or more) technologies, such as WiFi APs, RFID, video based localization, bluetooth, and ultra-wide band, and
(b) features such as time and angle of arrival of a signal, signal strength, and trilateration~\cite{yang2014three,luo2016pallas,xu2013scpl,musa2012tracking,li2015passive}.  
Such techniques can broadly be classified as either active or passive. 
Active approaches~\cite{deak2012survey,priyantha2000cricket} require individuals to download specialized software/apps and send information to a localization system~\cite{deak2012survey} which significantly limits technology adoption.
Non-participation and resistance to adoption renders applications that perform aggregate level analysis (e.g., analysis of space utilization and crowd flow patterns)  difficult to realize. Passive localization mechanisms, e.g.,~\cite{luo2016pallas,youssef2007challenges,xu2013scpl,seifeldin2012nuzzer,musa2012tracking,want1992active,li2015passive,ren2017d} address some of these concerns, but typically require  expensive \textit{external hardware}, significant parameter tuning that in turn requires ground truth data, and/or use APs in a monitor mode (in which case the AP cannot be used for data transmission and becomes a dedicated hardware for location determination). Tradeoffs to deal with such issues can cause limited precision, and are often not robust to dynamic situations such as movement of people, congestion, signal interference, and occlusion~\cite{ren2017d}.  Furthermore, techniques that offer high precision (e.g., ultra wide band) have significant cost and are not widely deployed. The semantic localization studied in this paper complements such indoor localization techniques with the goal of supporting  smart space applications that require associating individuals with semantically meaningful geographical spaces.


\vspace{-0.7em}
\section{Conclusions }
\vspace{-0.2em}
\label{sec:conclusion}

In this paper, we propose LOCATER that cleans existing WiFi connectivity datasets to perform semantic localization of individuals. The key benefit of LOCATER is that it: 1) Leverages existing WiFi infrastructure without requiring deployment of any additional hardware (such as monitors typically used in passive localization); 2) Does not require explicit cooperation of people (like active indoor localization approaches).
Instead, LOCATER leverages historical connectivity data to resolve coarse and fine locations of devices by cleaning connectivity data. Our experiments on both real and synthetic data show the effectiveness and scalability of LOCATER.  Optimizations made LOCATER achieve near real-time response.

LOCATER's usage of WiFi events, even though it does not capture any new data other than what WiFi networks already capture, still raises privacy concerns since such data is used for a purpose other than providing networking. Privacy concerns that arise and mechanisms to mitigate them, are outside the scope of this work and are discussed in~\cite{chen2017pegasus,ghayyur2018iot, panwar2019iot}. 
For deployments of LOCATER, we advocate to perform data collection based on informed consent allowing people to opt-out of location services if they choose to. 

\vspace{-0.5em}
\section{Appendix}
\label{sec:appendix}
\vspace{-0.2em}

\noindent{\bf Parameters Computation in Coarse Localization.}
\label{subsec:parameter-computation}
If ground truth data is available, we can use cross-validation to tune $\vLowBuild$ and $\vHighBuild$. Alternatively, we can estimate these parameters using the WiFi connectivity data as follows. For each device $d_i$, we count its average connection time to a WiFi AP (time difference between two consecutive connectivity events of $d_i$) based on a large sample of its connectivity data. Then, we plot a histogram where x-axis represents the duration and y-axis is the percentage of devices with a given duration between consecutive connections. The given frequency distribution can be approximated as a normal distribution  $\mathcal{N}$. We compute the confidence interval $(CI_l,CI_r)$ of the mean of $\mathcal{N}$ with $95\%$ confidence level, and set $\vLowBuild=CI_l$, $\vHighBuild=CI_r$. Intuitively, there is a $95\%$ probability that the mean of average duration of devices will fall in $(CI_l,CI_r)$, and the duration on the left side ($<=CI_l$) indicates that the device is inside the building while duration in the right side ($>=CI_r$) is outside.


\noindent {\bf Proofs for Section~\ref{subsec:alg}.}
\label{subsec:probability}
\noindent\textsc{Proof of Theorem~\ref{theory:max}}
Consider another possible world $W_0$ where some unseen devices are not in $r_j$. We denote by $W_{0}(d)$ the room where $d$ is located in $W_0$. We can transform $W$ to $W_0$ step by step, where in each step for a device that is not in $r_j$ in $W_0$, we change its room location from $r_j$ to $W_{0}(d)$. Assuming the transformation steps are $W,$ $W_n, . . . ,$ $W_1$, $W_0$, we can prove easily: $Pr(r_j|\bar{D}_n,W)>Pr(r_j|\bar{D}_n,W_n)>...>Pr(r_j|\bar{D}_n,W_1)>Pr(r_j|\bar{D}_n,W_0)$. 

Theorem 2 can be proven using a similar approach. The proof is included in the extended version of the paper~\cite{locater2020}.


\noindent\textsc{Proof of Theorem~\ref{theory:expected}}
We compute each possible world's probability based on the probabilities of the rooms being the answer, which are computed based on observations on $\bar{D}_n$. 

\vspace{-0.5em}
{\scriptsize	
	\begin{align}
	 expPr(r_j|\bar{D}_n) &=   \sum\limits_{W \in \mathcal{W}(D_n\backslash \bar{D}_n)}Pr(W)Pr(r_j | \bar{D}_n,W) \\ \nonumber 
	 &= \sum\limits_{W \in \mathcal{W}(D_n\backslash \bar{D}_n)}Pr(W | \bar{D}_n)\frac{Pr(r_j,\bar{D}_n,W)}{Pr(\bar{D}_n,W)} \\ \nonumber
	 &= \sum\limits_{W \in \mathcal{W}(D_n\backslash \bar{D}_n)}Pr(W | \bar{D}_n)\frac{Pr(\bar{D}_n)Pr(r_j,W| \bar{D}_n)}{Pr(\bar{D}_n)Pr(W|\bar{D}_n)} \\ \nonumber
	 &= \sum\limits_{W \in \mathcal{W}(D_n\backslash \bar{D}_n)}Pr(W | \bar{D}_n)\frac{Pr(\bar{D}_n)Pr(r_j| \bar{D}_n)Pr(W| \bar{D}_n)}{Pr(\bar{D}_n)Pr(W|\bar{D}_n)} \\ \nonumber
	 &= \sum\limits_{W \in \mathcal{W}(D_n\backslash \bar{D}_n)}Pr(W| \bar{D}_n)Pr(r_j | \bar{D}_n) \\ \nonumber
	 &=Pr(r_j| \bar{D}_n)
	\end{align}
}
\vspace{-2em}
{
\small
\begin{acks}
This material is based on research sponsored by HPI and DARPA under Agreement No. FA8750-16-2-0021. 
The U.S. Government is authorized to reproduce and distribute reprints for Governmental purposes notwithstanding any copyright notation thereon. The views and conclusions contained herein are those of the authors and should not be interpreted as necessarily representing the official policies or endorsements, either expressed or implied, of DARPA or the U.S. Government. 
This work is partially supported by NSF Grants No. 1527536,  1545071, 2032525, 1952247, 1528995 and 2008993.
\end{acks}
}
\balance
\bibliographystyle{ACM-Reference-Format}
\bibliography{refs}

\end{document}